\documentclass[iop,english,twocolappendix,numberedappendix,showpacs,superscriptaddress,appendixfloats,tighten,apj, twocolumn]{aastex6}
\usepackage{amsmath,amssymb,color,latexsym,natbib,graphicx}
\usepackage{dcolumn,ulem}
\usepackage{bm}% bold math
\usepackage{float}
\usepackage{hyperref}
\usepackage{amstext}
\usepackage{graphicx}
\usepackage{babel}
\usepackage{hyperref}
\usepackage{color}
\def\bb{{\bf B}}
\def\vv{{\bf v}}
\def\vA{{\bf v_A}}
\def\ww{{\bf w}}

\def\xx{{\bf x}}
\def\zz{{\bf z^\pm}}
\def\zzp{{\bf z^+}}
\def\zzm{{\bf z^-}}
\def\be{\begin{equation}}
\def\ee{\end{equation}}
\def\ba{\begin{eqnarray}}
\def\ea{\end{eqnarray}}
\def \pmbmath{\mathpalette\pmbmathaux}
\def \pmbmathaux#1#2{
         \pmbtext{$#1#2$}}
\def \pmbtext#1{\leavevmode
     \setbox0\hbox{#1}
     \kern0,4pt \copy0 \kern-\wd0
     \kern-0,2pt \raise0,3pt \box0 }

\begin{document}
\title{Energy cascade rate in compressible fast and slow solar wind turbulence}
%\title{Energy cascade rate in compressible MHD turbulence: a comparative study in the fast and slow solar wind}
\author{L.\,Z. Hadid\altaffilmark{1}, F. Sahraoui\altaffilmark{1} and S. Galtier\altaffilmark{1,2}}
\email{lina.hadid@lpp.polytechnique.fr}
\affil{$^1$ LPP, CNRS, Ecole polytechnique, UPMC Univ Paris 06, Univ. Paris-Sud, Observatoire de Paris, Universit\'e Paris-Saclay, Sorbonne Universit\'es, PSL Research University, 91128 Palaiseau, France}

\affil{$^2$ D\'epartement de Physique, Universit\'e Paris-Sud, Orsay, France}

\begin{abstract}   
Estimation of the energy cascade rate in the inertial range of solar wind turbulence has been done so far mostly within the incompressible magnetohydrodynamics (MHD) theory. Here, we go beyond that approximation to include plasma compressibility using a reduced form of a recently derived exact law for compressible, isothermal MHD turbulence. Using in-situ data from the THEMIS/ARTEMIS spacecraft in the fast and slow solar wind, we investigate in detail the role of the compressible fluctuations in modifying the energy cascade rate with respect to the prediction of the incompressible MHD model. 
In particular, we found that the energy cascade rate: i) is amplified particularly in the slow solar wind; ii) exhibits weaker fluctuations in spatial scales, which leads to a broader inertial range than the previous reported ones; iii) has a power law scaling with the turbulent Mach number; iv) has a lower level of spatial anisotropy.
Other features of solar wind turbulence are discussed along with their comparison with previous studies that used incompressible or heuristic (non exact) compressible MHD models. 
\end{abstract}

\maketitle

%----------------------------------------------------
\section{Introduction}\label{intro}
%------------------------------------------------------
A longstanding problem in the solar wind is its non-adiabatic cooling. This is manifested by the observations that the solar wind proton temperature decreases slowly as function of the radial distance from the Sun in comparison to the prediction of the adiabatic expansion model of the solar wind \citep{Marsch,Vasquez07}. Several scanarii have been proposed to explain that observations, e.g. pick up ions \citep{Matthaeus99,Smith01,Isenberg03,Marsch06}. The candidate that has driven much efforts is certainly the local heating of the solar wind plasma via turbulence~\citep{Bruno05,jltp,Smith06}. Large scale (MHD) turbulence can indeed serve as a reservoir of energy that cascades down to the small (kinetic) scales where it can be dissipated by some kinetic effects, which remain to be elucidated \citep{Goldstein94,Leamon98, Sahraoui09,Sahraoui10}. The underlying assumption is that all the energy that is injected at some large scale in the solar wind will cascade within the inertial range without dissipation, until it reaches the ion scale where it is eventually converted into thermal (heating) or kinetic (acceleration) energy of the plasma particles. This has led to intensive research work aiming at estimating the energy cascade rate in the solar wind using in-situ spacecraft data. A direct evidence of the presence of an inertial energy cascade in the solar wind was obtained using the so-called Yaglom law \citep{macbride05,luca,macbride08,Marino08,Marino11}. 
It is a universal law derived analytically from the incompressible MHD equations \citep{PP98a} (hereafter PP98) under the assumptions of homogeneity, stationarity, isotropy of the turbulent fluctuations and in the asymptotic limit of large kinetic and magnetic Reynolds numbers. Another fundamental assumption in those works is that compressible fluctuations play a minor role in the turbulent cascade. A first attempt to include the compressibility in estimating the energy cascade rate was made in \cite{Carbone09} (hereafter C09) using heuristic arguments. Indeed, a modified form of the Els\"asser variables was introduced which considered the local (instead of the mean) plasma density and a new density-weighted velocity 
\begin{equation}\label{eq_ww}
{\bf w}^\pm = \rho^{1/3}\Big{(}\vv \pm \frac{{\bf B}}{\sqrt{\rho\mu_0})}\Big{)} ,
\end{equation}
with $\rho$ is the density, $\vv$ the velocity, ${\bf B}$ the magnetic field and $\mu_0$ the permeability of free space. This form was inspired by the work of \cite{kritsuk07} (see also \cite{Schmidt08}) who showed numerically, in the context of supersonic interstellar turbulence, that the density-weighted velocity offers a better understanding of isothermal compressible hydrodynamic turbulence. 
The application of C09 to the fast solar wind data showed a better scaling relation of the energy flux than with PP98~\citep{Carbone09}. Furthermore, a significant increase of the turbulent cascade rate was evidenced and was shown to be sufficient to account for the local heating of the non-adiabatic solar wind expansion. 

A first attempt to include the compressible fluctuations of the solar wind in the turbulence cascade using a more rigorous approach has been done recently~\citep{Banerjee16}. In that paper, an exact law derived for compressible isothermal turbulence by~\cite{Banerjee13} (hereafter BG13) was used as well as the in-situ fast solar wind data measured by the THEMIS B/ARTEMIS P1 spacecraft \citep{Auster09,McFadden09}. Two important improvements in the estimation of the energy cascade rate using the BG13 model were obtained: i) a broader inertial range that extended for more than two decades of scales; ii) a higher energy cascade rate (up to $3$--$4$ times) than the estimation from the PP98 model. However, two discrepancies with the results of \cite{Carbone09} were found. First, the amplification of the cascade rate is smaller than that obtained in C09. Second, the origin of that enhancement is due to the new compressible terms in the BG13 model and not to the compressible Yaglom term~\citep{Banerjee16}. In the present study we extend the application of the BG13 model to a larger statistical sample and address new questions related to the differences between the slow and fast solar wind (known to have different levels of compressibility and different correlations between the magnetic and the velocity fields), the nature (direct versus inverse) of the turbulent cascade and the role of the cross-helicity, the effect of the turbulent Mach number and the plasma compressibility on the spatial anisotropy of the cascade rate. Throughout the paper, systematic comparisons with the incompressible model are made to highlight the role of the plasma compressibility. Discrepancies with the C09 model will be eventually discussed.

The manuscript is structured as follows: in Section~\ref{models} we recall the basic equations and assumptions of the different theoretical models used in this work, in Section~\ref{data} we describe the procedure we used to select our data samples, in Section~\ref{results} we present the main results of the study along with their comparisons to previous works, in Section~\ref{discussion} we discuss the origin of the discrepancies with the results reported in~\cite{Carbone09} and other caveats related to the theoretical models used and the data selection and, eventually, in Section~\ref{conclusion} we provide a summary of the results.
%-----------------------------------------------------------------------
\section{THEORETICAL MODELS} \label{models} 
%-----------------------------------------------------------------------

We briefly recall the basic equations of the three theoretical models, namely PP98, C09 and BG13, that will be used throughout the paper. 
These models are based on MHD which is a relevant model for most of the astrophysical plasmas \citep{G16}.  
%%%%%
%%%%%
\paragraph*{{\bf Incompressible model:}}  The PP98 law is written in terms of the Els\"asser variables $\zz = \vv \pm \vA$, where $\vv$ is the plasma flow velocity, $\vA \equiv \bb/\sqrt{\mu_0 \rho_0}$ is the magnetic field normalized to a velocity and $\rho_0=\langle \rho \rangle$ is the mean plasma density. It reads in the isotropic case
\be
- \frac{4}{3} \varepsilon_{I} \ell = 
\left\langle {\left(\delta \zzp \right)^2 \over 2} \delta z_{\ell}^- + {\left(\delta \zzm \right)^2 \over 2} \delta z_{\ell}^+ \right\rangle \rho_0 
\equiv {\cal F}_I (\ell) \, , \label{pp98a}
\ee
where the general definition of an increment of a variable $\psi$ is used, i.e. $\delta \psi \equiv \psi (\xx + \boldsymbol{\ell}) - \psi (\xx)$. 
The longitudinal components are denoted by the index $\ell$ with $\ell \equiv \vert \boldsymbol{\ell} \vert $, $\langle \cdot \rangle$ stands for the statistical average and $\varepsilon_{I}$ is the dissipation rate of the total energy. Note that in S.I. units, we have the relation $\rho_0= 1.673 \times 10^{-21}  \left\langle n_p \right\rangle $. 

%%%%%
\paragraph*{\bf{Heuristic compressible model}:} The heuristic C09 law is built from expression (\ref{pp98a}). The Els\"asser variables are simply replaced by a cube-root density weighted compressible Els\"asser variables $\ww^\pm \equiv \rho^{1/3} \zz$. Then, the isotropic law becomes 
\be
- \frac{4}{3}  \varepsilon_W \ell = 
\left\langle {\left( \delta \ww^+ \right)^2 \over 2} \delta w_{\ell}^- + {\left(\delta \ww^- \right)^2 \over 2} \delta w_{\ell}^+ \right\rangle 
\equiv {\cal F}_W (\ell) \, , \label{C09relation}
\ee
where $\varepsilon_W$ is the dissipation rate of the total compressible energy (following the notations introduced by \cite{Carbone09} that means 
$2\varepsilon_W = \varepsilon^+ + \varepsilon^-$).
Note that the renormalization proposed is inspired directly by studies of supersonic hydrodynamic interstellar turbulence \citep{kritsuk07,Schmidt08}. 

%%%%%
\paragraph*{{\bf Compressible model}} Following the approach used in \cite{Banerjee16}, the original equations of the BG13 model can be reduced to the following compact form in the isotropic case
\be
-{4 \over 3} \varepsilon_C \ell = {\cal F}_{C+\Phi}(\ell) \, , \label{bg13reduced}
\ee
where 
\be
{\cal F}_{C+\Phi}(\ell) = {\cal F}_{1}(\ell) + {\cal F}_{2}(\ell)+{\cal F}_{3}(\ell) \, , 
\ee
and
\ba
{\cal F}_{1}(\ell) &=&\left\langle \frac{1}{2} \left[ \delta ( \rho \mathbf{z}^-) \cdot \delta \mathbf{z}^-  \right]  {\delta  {z}_{\ell}^+}
+  \frac{1}{2} \left[  \delta ( \rho  \mathbf{z}^+) \cdot \delta \mathbf{z}^+ \right]  {\delta {z}_{\ell}^-} \right\rangle \, , \nonumber \\
{\cal F}_{2}(\ell)&=& \left\langle 2  \delta \rho \delta e  \delta v_{\ell} \right\rangle \, , \nonumber \\
{\cal F}_{3}(\ell)&=& \left\langle 2 {\overline{\delta} \left[ \left(1 + \frac{1}{\beta} \right) e + { v_A^2  \over 2}\right] \delta ( \rho_1 v_{\ell})}  \right\rangle \, ,
\label{fcphi}
\ea
where by definition $\overline{\delta} \psi \equiv (\psi (\xx + \boldsymbol{\ell}) + \psi (\xx))/2$,  
$e=c_s^2 \ln (\rho /\rho_0)$ is the internal energy, with $c_s$ the constant isothermal sound speed, $\rho$ the local plasma density ($\rho = \rho_0 + \rho_1$) and $\beta = 2 c_s^2 / v_A^2$ is the local ratio of the total thermal to magnetic pressure ($\beta=\beta_e+\beta_p$). We recall that, contrary to incompressible MHD theory, the BG13 compressible model yields an energy cascade rate that is not simply related to third order expressions of different turbulent fluctuations, but rather involves more complex combinations of the turbulent fields in the new flux and source terms.

To obtain Equations~(\ref{bg13reduced})--(\ref{fcphi}) several assumptions have been used (see details in~\cite{Banerjee16}). 
First, the source terms have been neglected based on the argument they are probably important only in supersonic turbulence whereas solar wind turbulence is subsonic~\citep{GB11,kritsuk13} and on preliminary estimation using numerical simulations of isothermal MHD turbulence~\citep{Servidio15}. Note that the source terms cannot be estimated reliably using single spacecraft data in this work because of the local spatial divergence involved in those terms. Second, the plasma $\beta$ is assumed to be nearly stationary, which is a stringent requirement in selecting the data to use in the present study. 
To these assumptions add up the classical ones generally used to derive similar equations in turbulence theories, namely statistical homogeneity and stationarity of the turbulent fluctuations. The statistical isotropic assumption is further made to obtain the reduced form given by Equations (\ref{bg13reduced})--(\ref{fcphi}) (this point will be discussed in more detail in Section~\ref{anisot}). In this work, it is these Equations~ that will be evaluated using spacecraft data in the fast and slow solar wind. 
%--------------------------------------------------------------------------------------------------------------
\section{DATA SELECTION}\label{data} 
%--------------------------------------------------------------------------------------------------------------
We used the THEMIS B/ARTEMIS P1 spacecraft data during time intervals when it was travelling in the free-streaming solar wind. The magnetic field data and plasma moments (density, velocity and temperature) were measured respectively by the Flux Gate Magnetometer (FGM) and the Electrostatic Analyzer (ESA). All data have 3 seconds time resolution (i.e., spin period). A large survey of the THEMIS/ARTEMIS data has been performed from the period 2008-2011 that covered both the fast and slow solar wind.  Fast winds are those having their average velocity  $ V> 450 \ \text{km} \, \text{s}^{-1}$. In selecting the data, we have tried to avoid intervals that contained significant disturbances or large scale gradients (e.g., coronal mass ejection or interplanetary shocks). As mentioned above, a limiting criterion in choosing the data is the condition of having a stationary plasma $\beta$, which has been checked for each case separately in this work. Another parameter that has been checked is the uniformity of $\Theta_{\bf VB}$, the angle between the local solar wind speed ${\bf V}$ and the magnetic field ${\bf B}$. Indeed, when using the Taylor hypothesis on single spacecraft measurements, the time sampling of the data is converted into a 1D spatial sampling of the turbulent fluctuations along the flow direction. Therefore, the stationarity of the angle $\Theta_{\bf VB}$ is required to guarantee that the spacecraft is sampling nearly the same direction of space with respect to the local magnetic field, which would ensure a better convergence in estimating the cascade rate. This point will be further developed in Section~\ref{dis_theta}. 

\begin{figure}
\includegraphics[width=1\columnwidth]{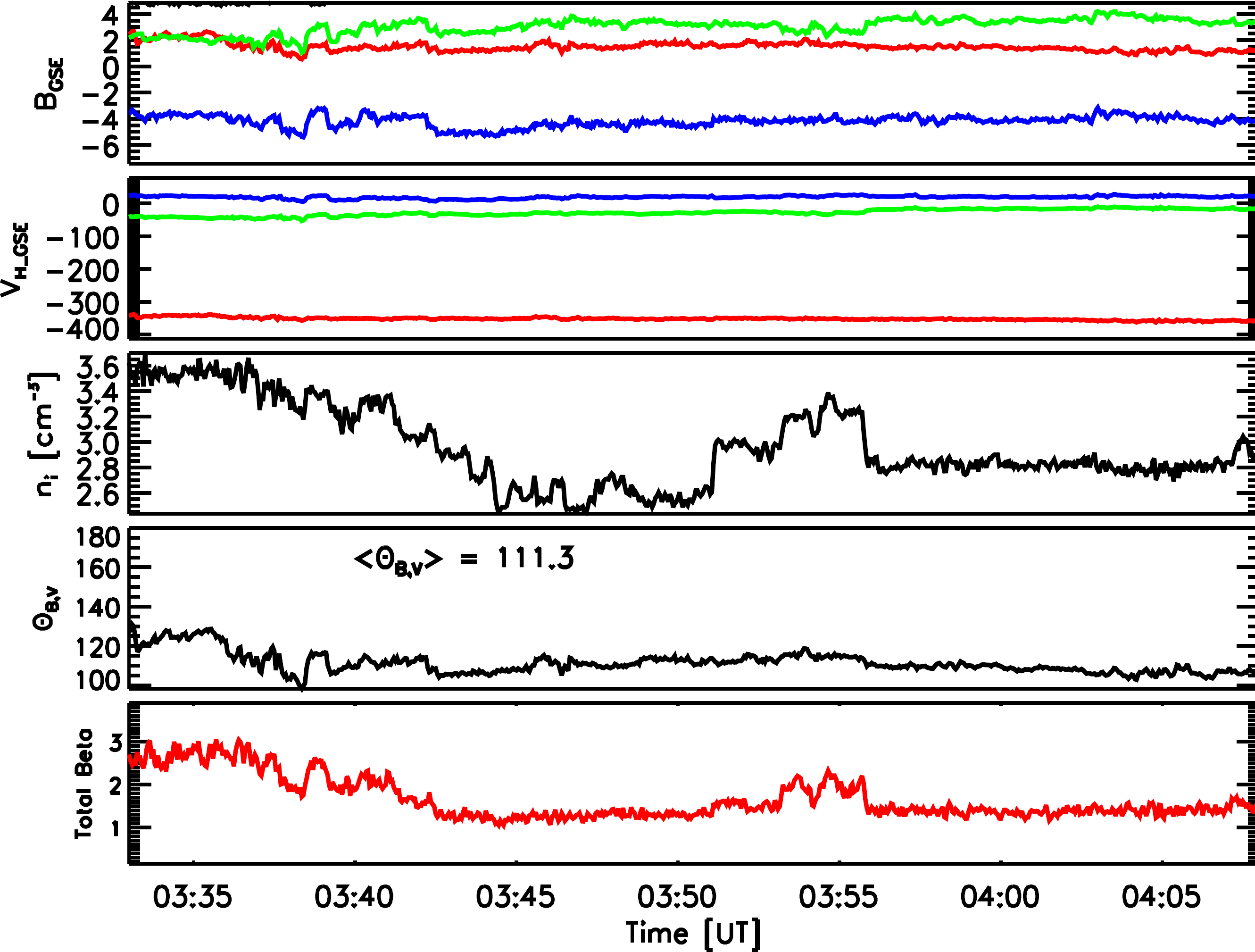}
\caption{From top to bottom: the solar wind magnetic field components (nT), ion velocity (km/s), ion number density, $\Theta_{\bf VB}$ angle 
and total plasma beta ($\beta=\beta_i+\beta_e$) measured by the FGM and ESA experiments onboard the THEMIS B spacecraft on 2009-11-20 from 03:33 to 04:08.}
\label{waveforms}  
\end{figure}

The obtained intervals that fulfilled all the previous criteria were divided into a series of samples of equal duration $\sim 35$mn, which
corresponds to a number of data points $N\sim 700$ with a $3$s time resolution. This number of points is larger than those used in previous studies based on ACE spacecraft data ($N\sim 150$ with intervals of $1$h and a time resolution of $24$s, see e.g., ~\cite{macbride08}). This allows for a more accurate estimation of the moments of the turbulent field increments. The duration of $35$mn ensures having at least one correlation time of the turbulent fluctuations estimated to vary in the range $\sim 20-30$mn. Eventually, the data selection yielded $148$ samples ($\sim 1\times10^5$ data points) in the fast solar wind and $182$ ($\sim 1.3\times10^5$ data points) in the slow solar wind. An example of the analyzed time intervals is shown in Figure~\ref{waveforms} in the slow solar wind.  

\begin{figure}
\includegraphics[width=1\columnwidth]{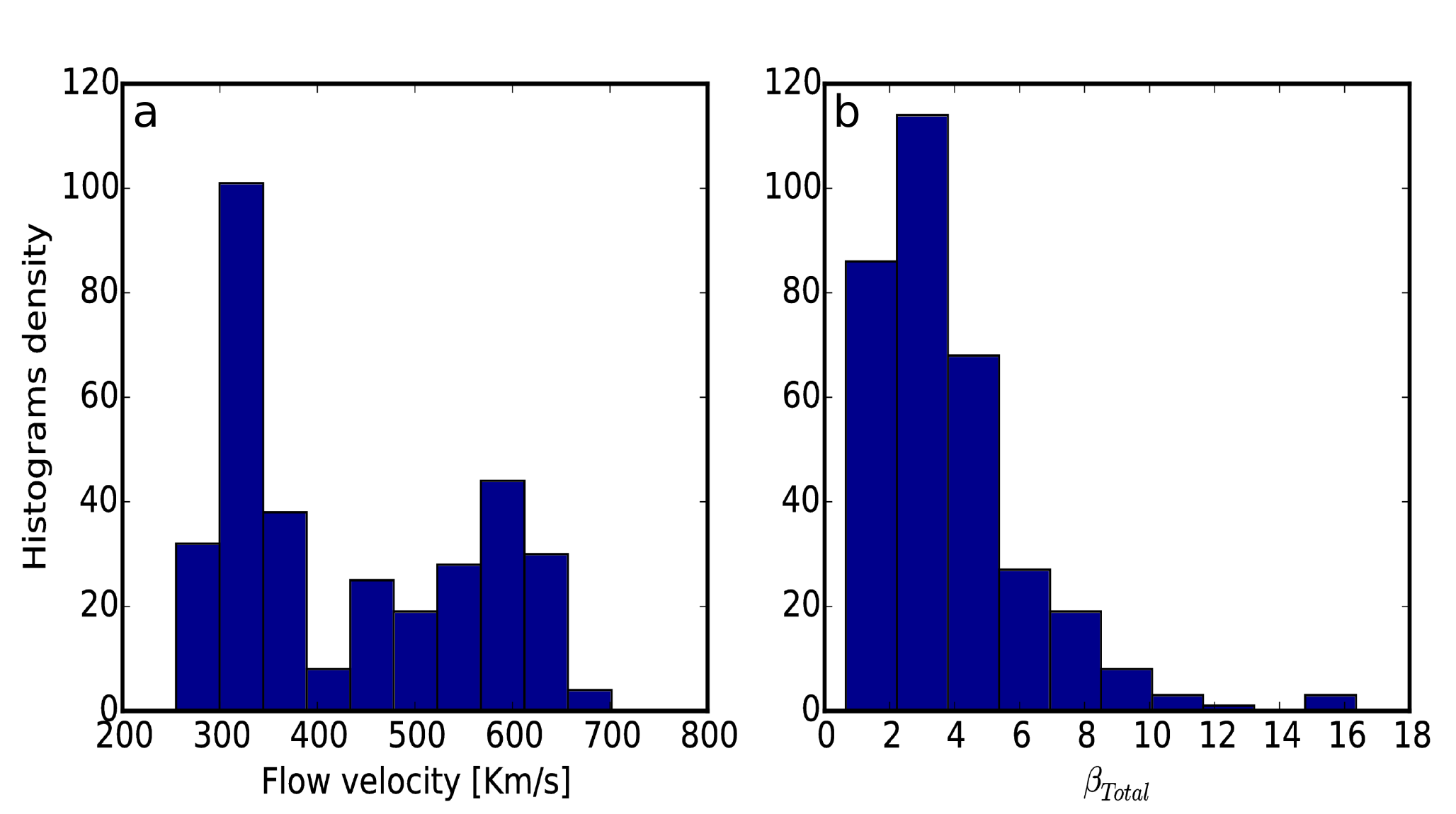}
\caption{The average solar wind speed (a) and the total plasma $\beta$ (b) for all the used data intervals.}
\label{Histo_parameters}  
\end{figure}

The average solar wind speed and plasma $\beta$ for all the statistical samples are shown in Figure~\ref{Histo_parameters}. Note that most of the values of $\beta$ are larger than 1.

%--------------------------------------------------------------------------------------------------------------
\section{OBSERVATIONAL RESULTS IN THE FAST AND SLOW SOLAR WIND}\label{results} 
%--------------------------------------------------------------------------------------------------------------

%-------------------------------------------------------------------------------------------------------
\subsection{Cascade rate versus plasma compressibility and turbulent Mach number} \label{firstsec}
%-------------------------------------------------------------------------------------------------------
For all the selected time intervals we computed the energy cascade rates $\varepsilon_I$ and $\varepsilon_C$ from the PP98 and the BG13 models using respectively Equation~(\ref{pp98a}) and Equations~(\ref{bg13reduced})--(\ref{fcphi}). To this end, we had constructed temporal structure functions of the different turbulent fields involved in those equations, namely ${\bf B}$, $n$ and $\vv$, at different time lags $\tau$. In order to probe into the scales of the inertial range, known to lie within the frequency range $\sim$[$10^{-4},1$]\,Hz (based on the observation of the Kolmogorov-like $-5/3$ magnetic energy spectrum~\citep{Bruno05}), we vary the time lag $\tau$ from 10\,s to 1000\,s thereby being well inside the targeted frequency range. Note that this range of scales is slightly shifted toward small scales in comparison to previous studied that used ACE data~\citep{macbride08}.  

\begin{figure}
\includegraphics[width=1\columnwidth]{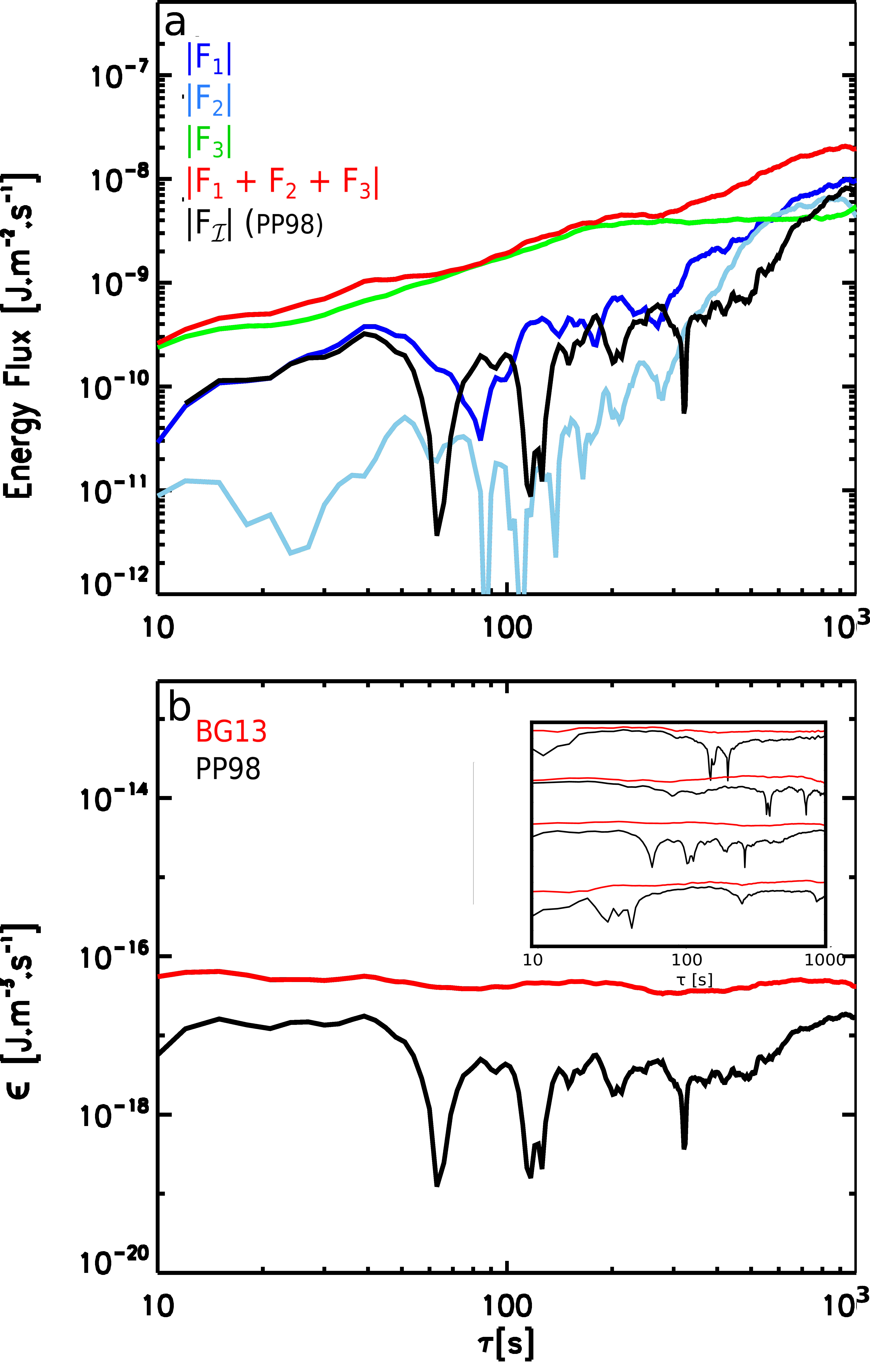}
\caption{(a) Comparison of the different fluxes $\vert {\cal F}_1 \vert$, $\vert {\cal F}_2 \vert$, $\vert {\cal F}_3 \vert$, ${\cal F}_{C+\Phi}$ and $\vert {\cal F}_I \vert$ (see text for the definitions) in the slow solar wind for the same event of Figure~\ref{waveforms}. (b) comparison between the corresponding turbulent cascade rates given by the PP98 and BG13 models. The compressibility is $\sim 11\%$.
The inset shows other examples for which the BG13 model gives a smoother cascade rate over two decades of scales than the PP98 model.}
\label{flux_example}  
\end{figure}

A detailed comparison of the different fluxes of the BG13 model is given in Figure~\ref{flux_example} (a). We see that the pure compressible flux
$\vert {\cal F}_3 \vert$ dominates the other fluxes for most of the time scales $\tau$. For comparison we also show the incompressible flux 
$\vert {\cal F}_I \vert$ given by the PP98 model which is clearly much lower than ${\cal F}_{C+\Phi}$. In Figure~\ref{flux_example} (b) are plotted the cascade rates 
deduced from the flux analysis. The estimate from BG13 gives a flat cascade rate over two decades of scales whereas the estimate from the PP98 exhibits 
hollows which are the manifestation of a change of sign. This difference is a generic behavior found in many other cases (see inset). The sign change of the incompressible cascade rate does not always occur at the same time lag $\tau$ as can be seen in Figure~\ref{flux_example}. This rules out the possible role of minor heavy ions, e.g. ${H_e}^{+}$, whose charateristic scales would belong to the range of scales analyzed in this work. This was confirmed by a visual check of the power spectra of the magnetic fluctuations which did not show any significant enhancement of power in the frequency range $[10^{-3},10^{-1}]$Hz, which would be caused by an energy injection via a kinetic plasma instability of heavy ions.

\begin{figure}
\includegraphics[width=1\columnwidth]{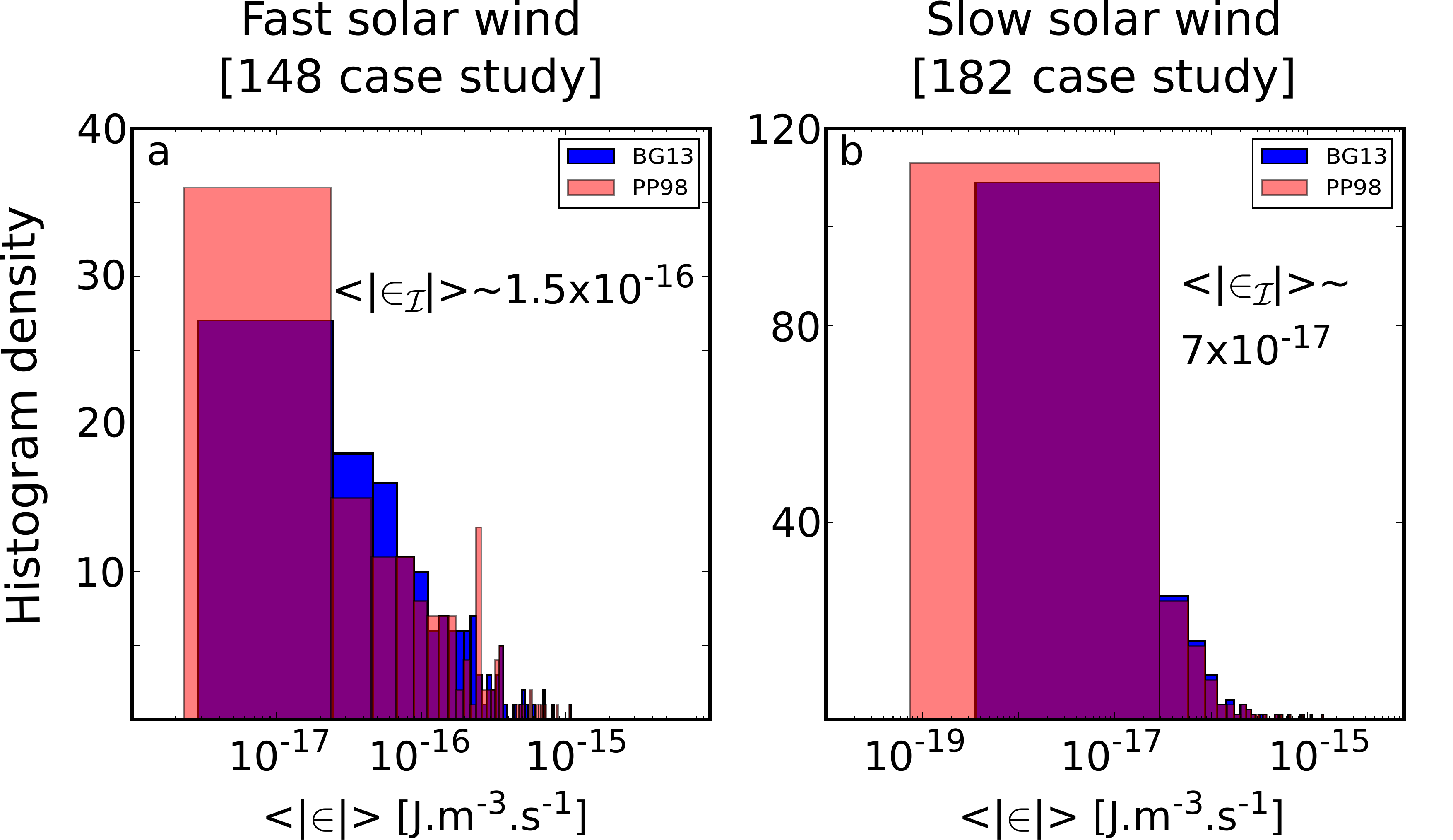}
\caption{Histograms of the absolute value of the cascade rates measured in the fast (a) and slow (b) solar wind. A systematic comparison is made 
between the incompressible and compressible predictions.}
\label{epsilon_histogam}  
\end{figure}

Examples of the obtained cascade rates computed in the fast and slow solar wind are shown in Figure~\ref{epsilon_histogam}.  
A first observation is that both the compressible and incompressible cascade rates $\langle|\varepsilon_C|\rangle$ and  $\langle|\varepsilon_I|\rangle$ are larger in the fast wind than in the slow wind as indicated by the histogram and the average (absolute) values. This confirms the previous finding regarding the incompressible cascade rate $\langle|\varepsilon_I|\rangle$ \citep{macbride08,Stawarz2009,Coburn2012} and shows that compressibility does not change that trend. 

\begin{figure}
\includegraphics[width=1\columnwidth]{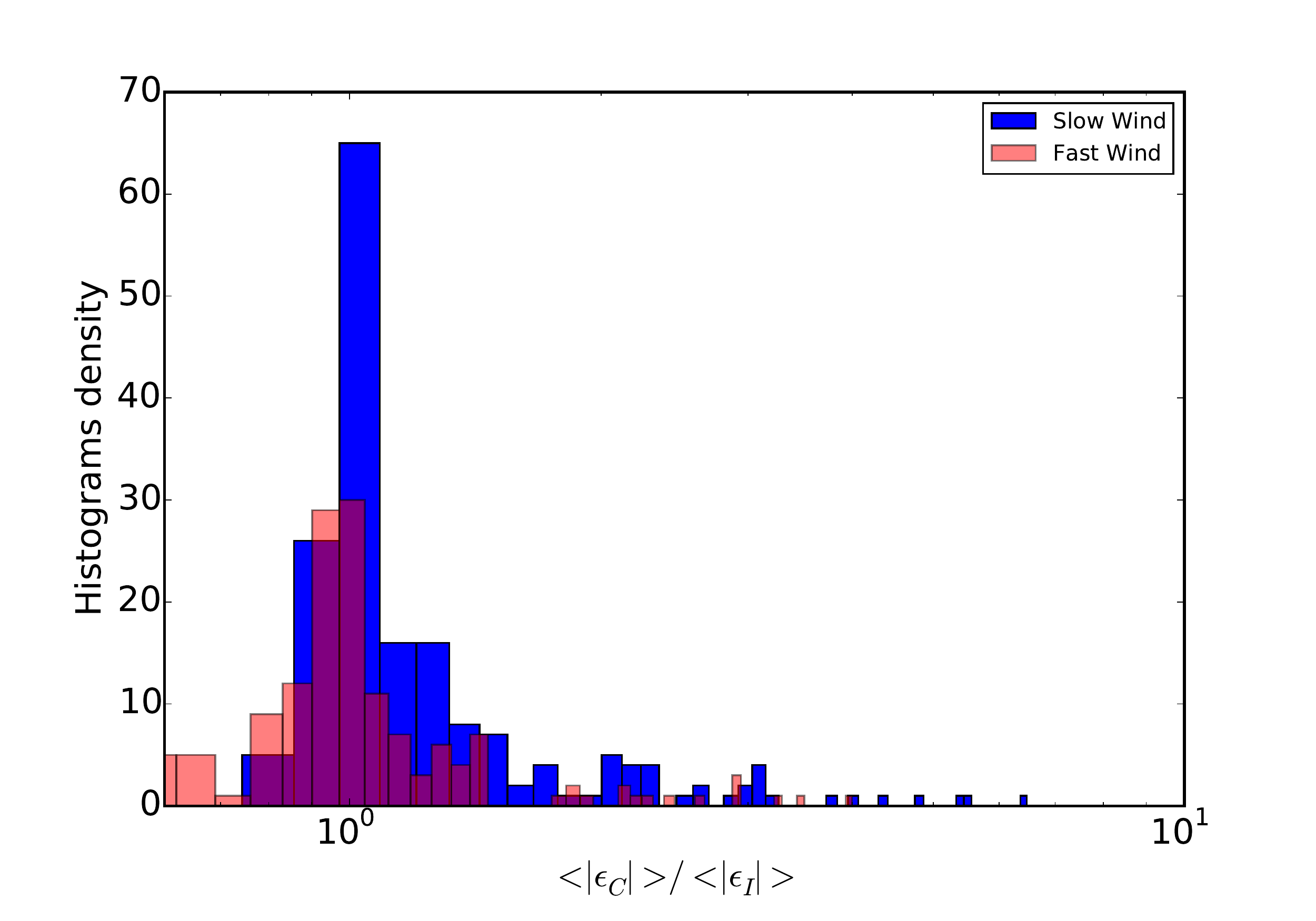}
\caption{Histograms of the ratio between the compressible to the incompressible cascade rate $R=\langle|\varepsilon_C|\rangle/\langle|\varepsilon_I|\rangle$ in the fast  (pink) and slow (blue) winds.}        
\label{histogram_ratio_BG13PP98}  
\end{figure}

In Figure~\ref{histogram_ratio_BG13PP98} we compare the ratio between the compressible to the incompressible cascade rate $R=\langle|\varepsilon_C|\rangle/\langle|\varepsilon_I|\rangle$ in the fast  and slow winds. Here we use the average value of the cascade rate over all the time lags $\tau$ within the range $10-1000\,s$. This may contrast with previous studies where the statistical results were given at a specific value of $\tau$ within the inertial range \citep{Podesta09,Coburn15}. Indeed, as it will be discussed in Section~\ref{dis_sign}, the cascade rate may change its sign within a single time interval for two (or more) different values of $\tau$ in the inertial range, which makes the choice of a single value of $\varepsilon$ at a given value of $\tau$ questionable. This motivated a new criterion applied to further narrow down the selection of our time intervals: we kept only those samples for which the compressible cascade rate shows a constant (negative or positive) sign for all time lags in the range $10-1000$ s. 

A first feature that can be seen in Figure~\ref{histogram_ratio_BG13PP98}, and already reported in reported in~\cite{Banerjee16} regarding the fast solar wind, is that the plasma compressibility, while in average may not modify significantly the cascade rate (since the bulk of the distribution of the ratio $R$ is centred around 1), in some cases it does nevertheless amplify it by a factor of $3-4$. This trend is enhanced in the slow wind where the (blue) histogram of $R$ in Figure~\ref{histogram_ratio_BG13PP98} is found to shift to higher values (up to $7-8$) and for a larger number of events than in the fast wind. Note however that these amplication values remain smaller than those reported in~\cite{Carbone09} as it will be discussed in Section~\ref{dis_flow}.

\begin{figure}
\includegraphics[width=1\columnwidth]{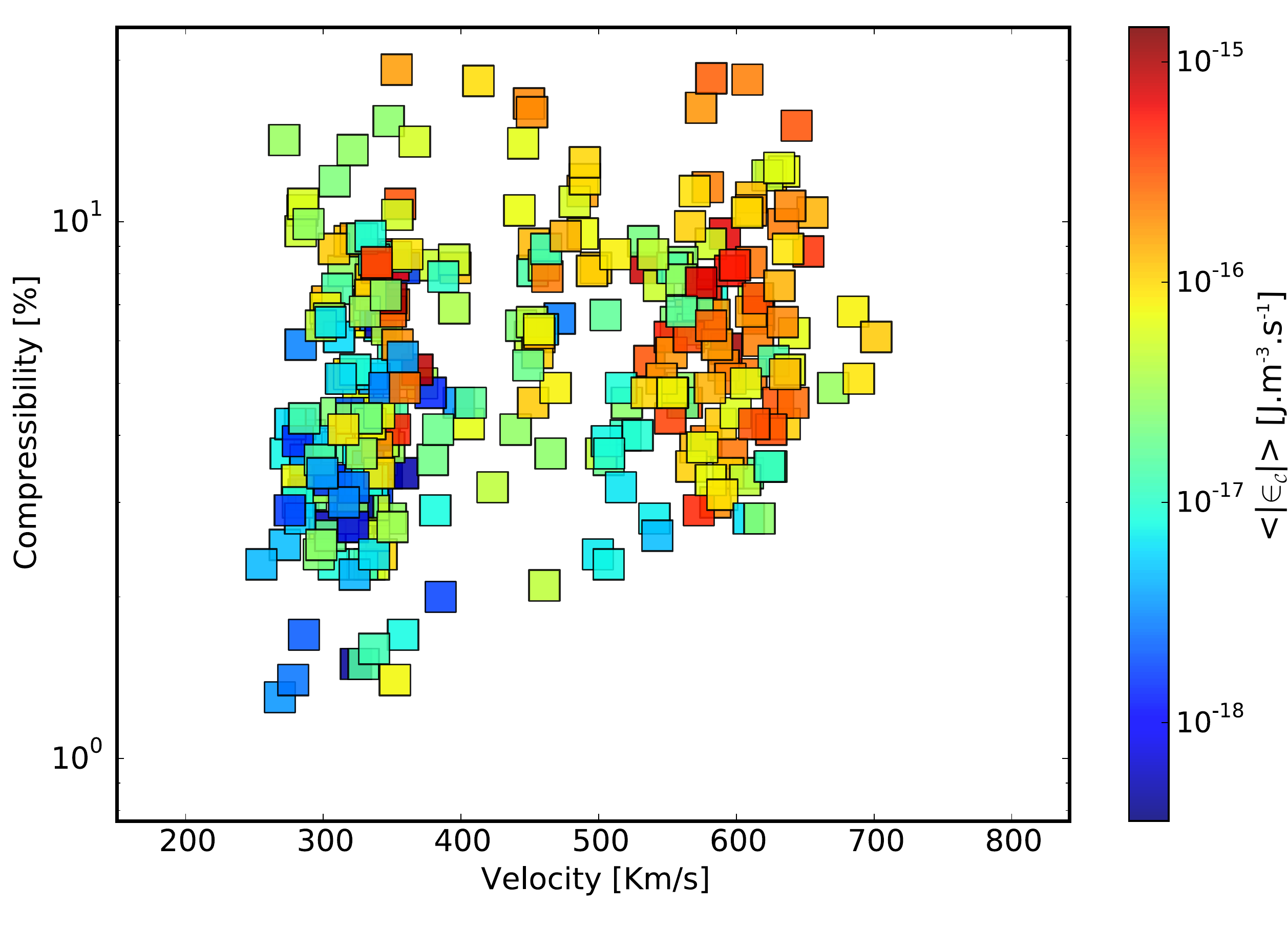}
\caption{Variation of the compressible cascade rate $\langle \vert \varepsilon_C \vert \rangle$ as function of the density variation and the wind speed.}
\label{2D_epsilon_velocity}  
\end{figure}

To evidence the role of the density fluctuations $\sqrt{\left( \langle \rho^2  \rangle - \langle \rho \rangle^2  \right)}/ \langle \rho \rangle$ in enhancing the cascade rate $\langle |\varepsilon_C|\rangle$ w.r.t. the incompressible one $\langle |\varepsilon_I|\rangle$ we plotted in Figure~\ref{2D_epsilon_velocity} $\langle|\varepsilon_C|\rangle$ as function of the wind speed and the density fluctuations. First, one can find the property discussed above that, overall, the fast wind has a higher $\langle |\varepsilon_C|\rangle$ than the slow wind. Moreover, one can see an increase in the cascade rate as compressibility increases in particular in the case of the slow wind. This trend is less evident in the case of the fast solar wind possibly because the spread in the compressibility values is smaller ($\sim 3\%-15\%$) than in the case of the slow wind ($\sim 1\%-20\%$).

\begin{figure}
\includegraphics[width=1\columnwidth]{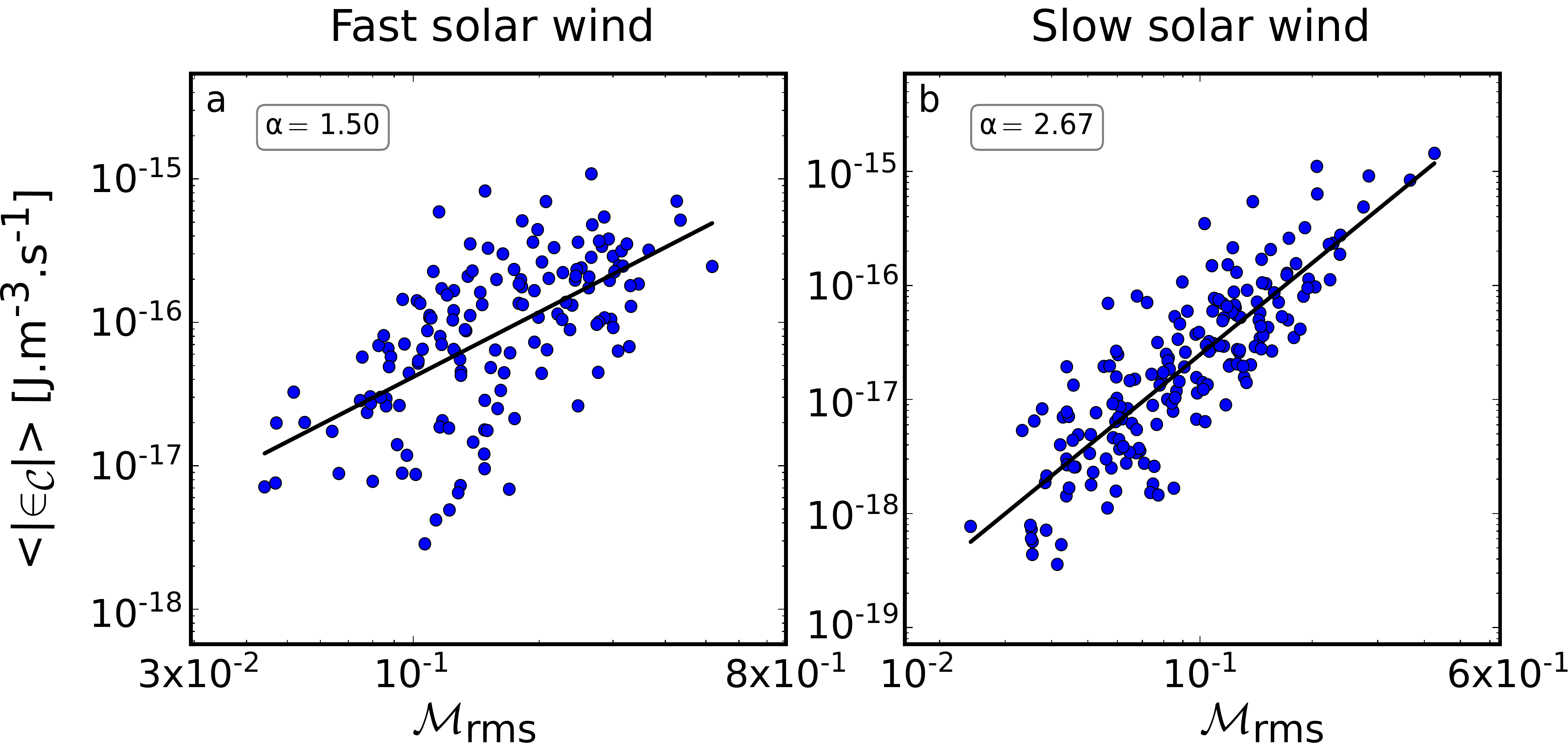}
\caption{Variation of the compressible cascade rate $\langle \vert \varepsilon_C \vert \rangle$ as function of the turbulent sonic Mach number $M_{rms}$ 
estimated in the fast (a) and the slow (b) solar wind.}
\label{2D_epsilon_mach}  
\end{figure}

The correlation is better seen with the estimated turbulent sonic Mach number defined as $M_{rms}=\sqrt{{v_1}^2/{c_s^2}}$ ($v_1$ being the average fluctuating plasma flow as shown in Figure~\ref{2D_epsilon_mach}. The slow wind shows a clear power-law in 

\begin{equation}
\varepsilon_C \sim M_{rms}^{2.67} ,
\end{equation}
while the fast wind exhibits more spread around an approximate power-law
\begin{equation}
\varepsilon_C \sim M_{rms}^{1.5}
\end{equation}

%-------------------------------------------------------------------------------------------------------
\subsection{Cascade rate versus the energy of the turbulent fluctuations} 
%-------------------------------------------------------------------------------------------------------
Another interesting feature that can be analyzed is the dependence of the cascade rate $\varepsilon_C$ on the energy of the compressible turbulent fluctuations $E^{comp}_1$ and the possible existence of a scaling law relating each of the energy components to $\varepsilon_C$. Indeed, unlike in the incompressible model PP98, the total energy of the fluctuations is not given simply by 

\begin{equation}\label{eq_Eincomp}
 E^{inc}_1=\frac{\rho_0}{4}( {\bf z_1^+}^2+ {\bf z_1^-}^2)=\frac{1}{2}\rho_0 v_1^2+\frac{1}{2\mu_0}B_1^2 , 
 \end{equation}
 
where $v_1$ and $B_1$ are the fluctuating velocity and magnetic fields, but includes the fluctuating internal energy $U_1$, hence
 
\begin{equation}\label{eq_Ecomp1}
E^{comp}_{1}=\frac{\rho_0}{4}({\bf z_1^+}^2+ {\bf z_1^-}^2)+ U_1 , 
\end{equation}

where, to the lowest order of $\rho_1/\rho_0$, $U_1$ can be written as
\begin{equation}\label{eq_Ecomp2}
U_1 = \rho_0c_s^2 \ln(1+\rho_1/\rho_0) .
\end{equation}

\begin{figure}
\includegraphics[width=1\columnwidth]{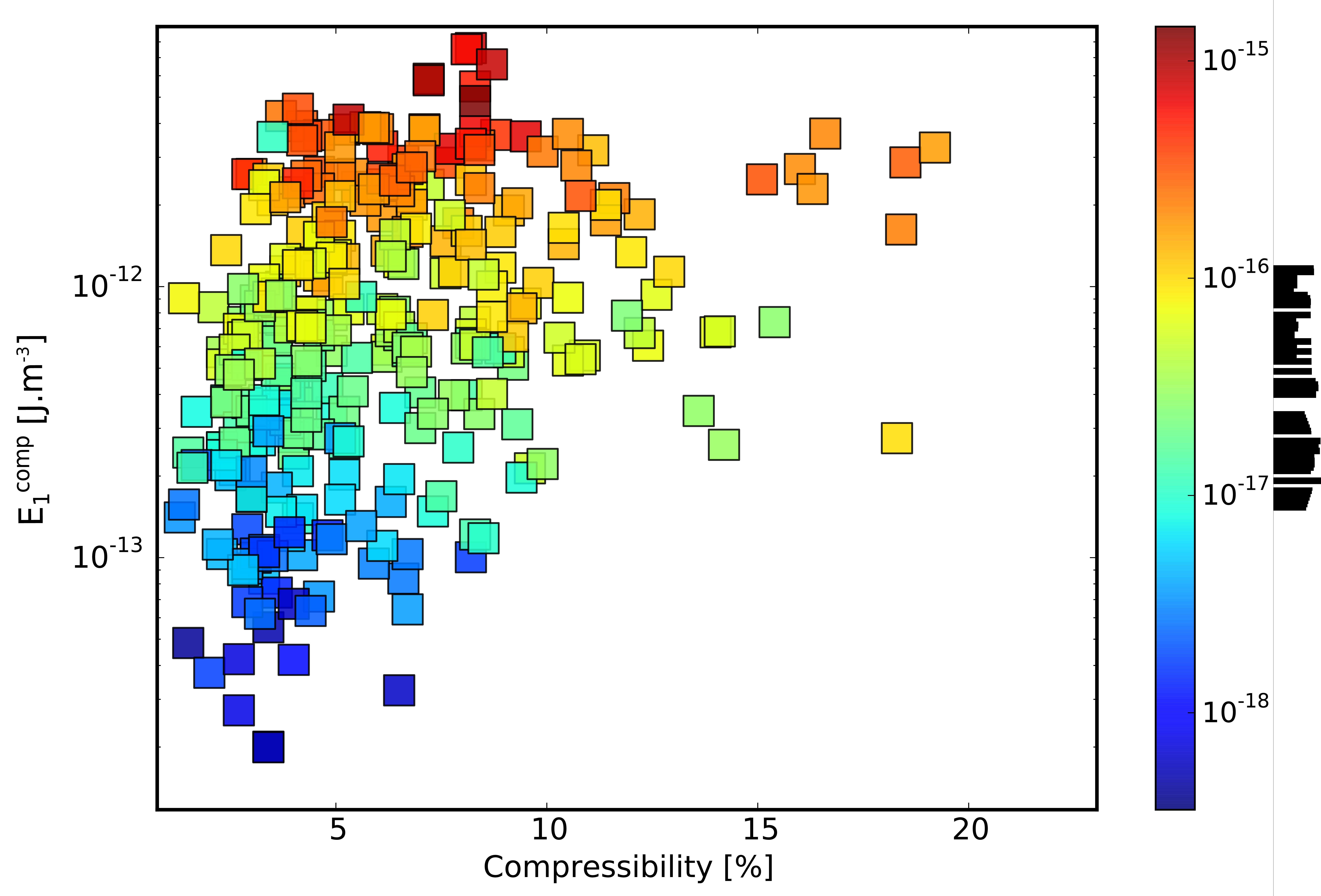}
\caption{The compressible cascade rate $\langle \vert \varepsilon_C \vert\rangle$ (in color) as function of the compressibility and the total energy of the turbulent fluctuations ($E_{1}^{comp}$).}
\label{2D_epsilon_energy}  
\end{figure}

First, we plotted in Figure~\ref{2D_epsilon_energy} the variation of $\langle |\varepsilon_C|\rangle $ as function the compressibility and the total energy of the turbulent fluctuations $E_1^{comp}$ given by Equations~(\ref{eq_Ecomp1})--(\ref{eq_Ecomp2}) computed in the fast and slow wind. One can see clearly that the higher is the amplitude of the fluctuation the larger is the cascade rate $\langle|\varepsilon_C|\rangle$. This observation si valid both in the fast and in the slow wind and is consistent with previous observations \citep{Smith06,macbride08}. Note that there is no significant variation of $\langle|\varepsilon_C|\rangle$ as function of compressibility at a fixed value of the energy of the turbulent fluctuations.

\begin{figure}
\includegraphics[width=1\columnwidth]{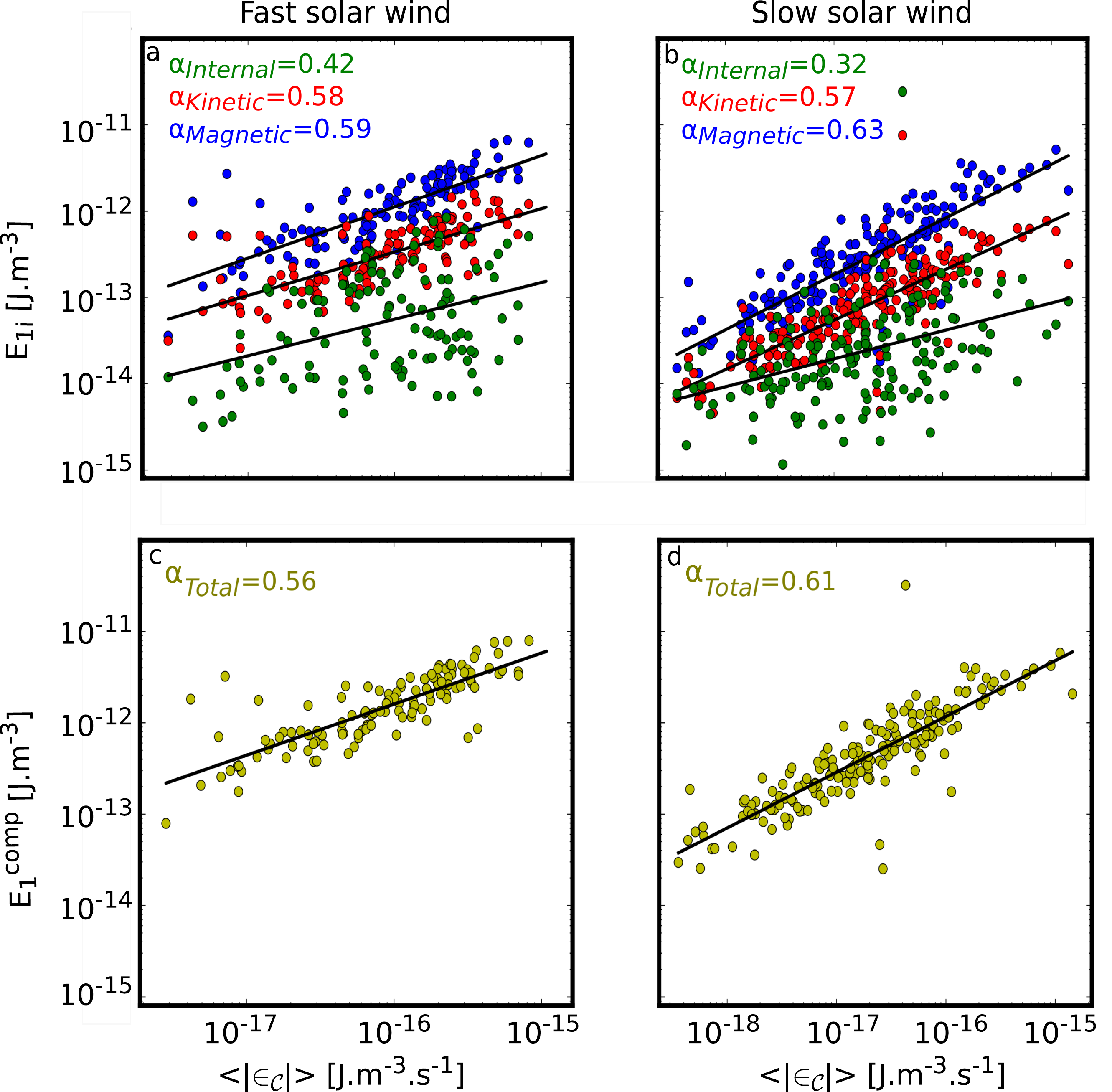}
\caption{ The magnetic ($E_{M}$, blue), kinetic ($E_{K}$, red) and internal ($E_{I}$, green) compressible energies plotted as a function of the compressible energy cascade rate $\langle|\varepsilon_C|\rangle$ in the fast (a) and slow (b) winds. (c-d) the total compressible energy $E_{1}$ as a function of $\langle|\varepsilon_C|\rangle$.}        
\label{epsilon_E}  
\end{figure}

Figure~\ref{epsilon_E} shows the three components of the total energy of the fluctuation $E^{comp}_1$ as function of 
the estimated compressible cascade rate $|\varepsilon_C|$  for all the statistical samples analyzed in the fast and slow solar wind. 
First, one can see that, statistically, the magnetic energy dominates over the kinetic and internal energies, 
the latter being the smallest, confirming the same results of \cite{Podesta2007}. Second, a relatively clear power-law scaling between $E_{1i}$ ($i=K,M,I$  for kinetic, magnetic and internal energies) and $|\varepsilon_C|$ can be evidenced with nearly the same slope in the fast and slow winds

  \begin{equation}\label{scaling1}
   E_{1K}\sim \varepsilon_C^{0.57} ,
  \end{equation}
  and 
  \begin{equation}\label{scaling1}
   E_{1M} \sim \varepsilon_C^{0.60} .
  \end{equation}
The scaling of the internal energy is shallower and is different for the two types of wind: 
   \begin{equation}\label{scaling2}
   E_{1I} \sim \varepsilon_C^{0.42} ,
  \end{equation}
for the fast wind, and
   \begin{equation}\label{scaling3}
   E_{1I} \sim \varepsilon_C^{0.32} .
  \end{equation}
  
for the slow wind. While the scaling of the magnetic and kinetic energies with the cascade rate are very close to the theoretical prediction from the Kolmogorov theory \citep{frisch}, $ E_1 \sim \varepsilon^{2/3}$, to the best of our knowledge, no theoretical prediction exists so far to help interpreting the empirical laws~(\ref{scaling2})--(\ref{scaling3}). 

%-------------------------------------------------------------------------------
\subsection{Role of the different flux terms} 
\begin{figure}
\includegraphics[width=1\columnwidth]{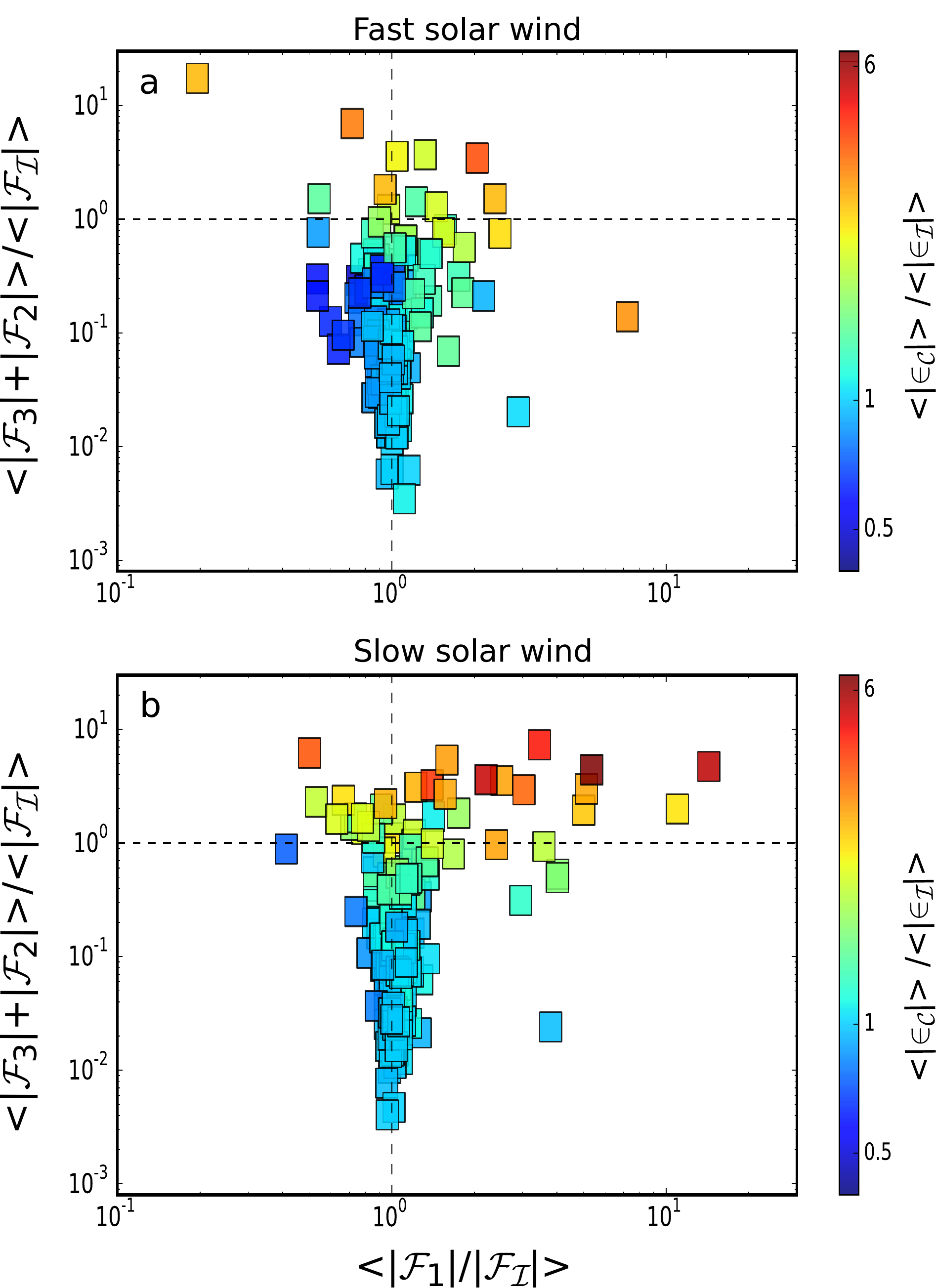}
\caption{Estimation of the contribution of the compressible fluxes w.r.t. incompressible (Yaglom) 
flux to the compressible cascade rate for the fast (a) and slow (b) winds.}
\label{epsilon_fluxterms}  
\end{figure}

%-------------------------------------------------------------------------------
To gain insight into the role of the different flux terms involved in estimating the compressible energy cascade rate, we plotted in Figure \ref{epsilon_fluxterms} statistical results about the contribution of the different compressible fluxes, ${\cal F}_1$, ${\cal F}_2$ and ${\cal F}_3$, relative to the incompressible (Yaglom) flux ${\cal F}_I$ for the slow and fast winds. A first observation is that most of the samples have their compressible Yaglom  
flux (${\cal F}_1$) of the order of the incompressible flux (${\cal F}_I$). This indicates that it is the new compressible fluxes ${\cal F}_2$ and ${\cal F}_3$ that contribute more to enhancing the compressible cascade rate $\varepsilon_C$ (w.r.t. $\varepsilon_I$) rather than the compressible Yaglom term ${\cal F}_1$. This is better seen when observing that high values of $\langle |\varepsilon_{C}| \rangle/ \langle |\varepsilon_{I}| \rangle$ (up to $\sim 4$ in the fast wind and up to $8$ in the slow wind) are observed when 
$(\langle |{\cal F}_2| \rangle + \langle |{\cal F}_3| \rangle)/\langle \vert {\cal F}_I \vert \rangle>1$. We recall that stronger amplification has been reported in~\cite{Carbone09}, which stems from an heuristic modification of the incompressible (Yaglom) term {\it via} density fluctuations. The discrepancy between that observation and the present ones will be enlightened in Section~\ref{dis_flow}. 
Note finally that the highest ratio $R = \langle \vert \varepsilon_{C}\vert  \rangle / \langle \vert \varepsilon_{I} \vert \rangle$ (i.e., highest amplification of the cascade rate due to compressible fluctuations) is observed in the top-right quarter (fast wind) of Figure~\ref{epsilon_fluxterms}, which corresponds to the cases when all the three terms ${\cal F}_1$, ${\cal F}_2$ and ${\cal F}_3$ dominate over the incompressible (Yaglom) term ${\cal F}_I$. The highest values of the ratio $R$ are also observed in this quarter for the slow wind.

%-------------------------------------------------------------------------------------
\subsection{Sign of the energy transfer rate and cross-helicity} \label{Sign} 
\begin{figure}
\includegraphics[width=1\columnwidth]{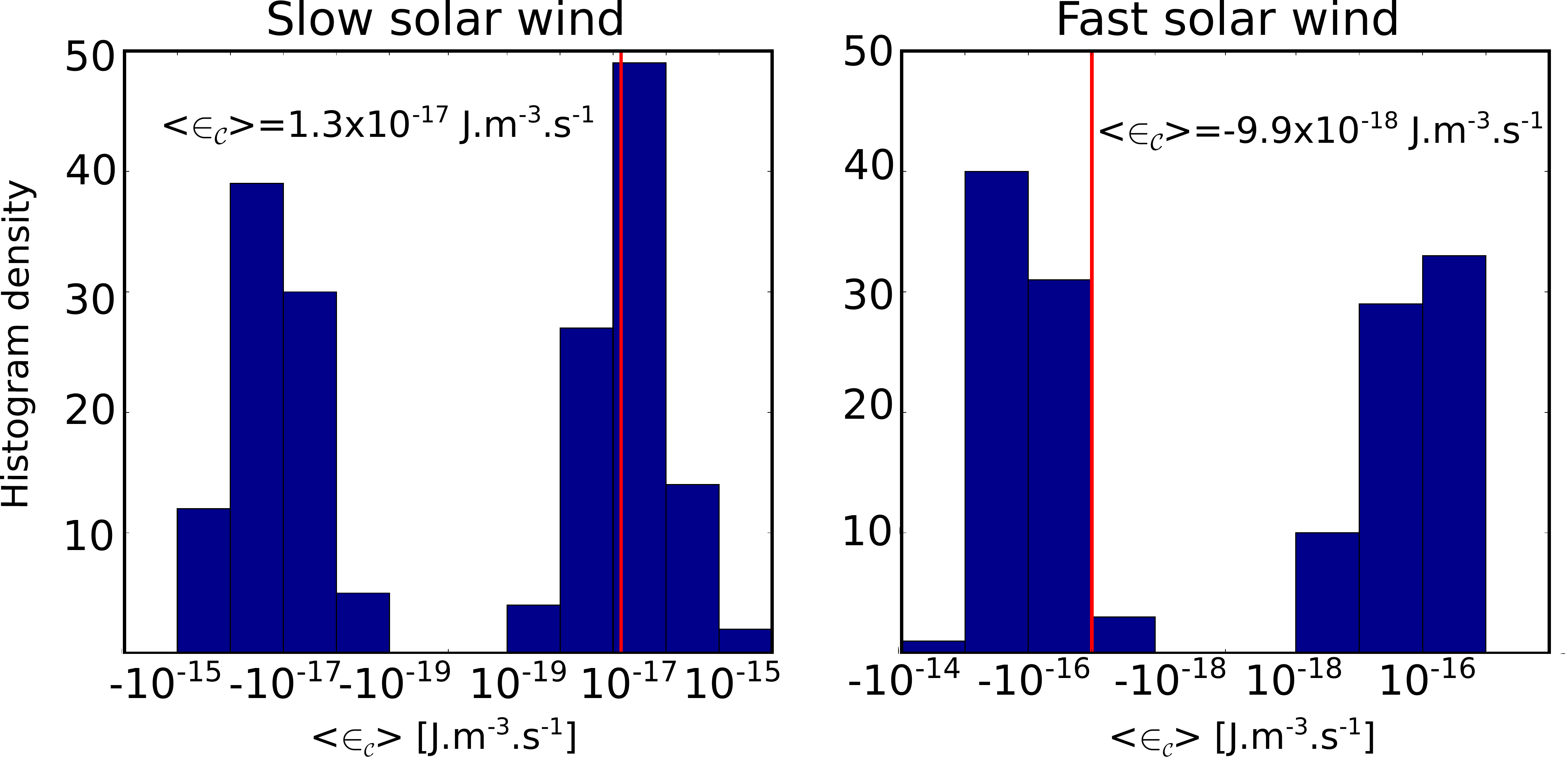}
\caption{Histograms of the signed energy cascade rate estimated using the compressible model BG13 in the fast (b) and slow (a) solar wind.}
\label{histogramsign_BG13}  
\end{figure}

%-------------------------------------------------------------------------------------
In this section we discuss the sign of the cascade rate as estimated from the incompressible (PP98) and compressible (BG13) models. We first recall that this property can be discussed only when the dependence of the energy flux on the time increments $\tau$ are converted into the spatial ones $l$ via the Taylor frozen-in flow assumption. With the positive convention of the time increments ($\tau>0$) used in this work, the Taylor hypothesis implies $l\sim -V\tau$. In this convention, positive (resp. negative) values of $\varepsilon_{I,C}$ correspond to a direct (inverse) energy cascade. 
The histograms of the signed compressible cascade rate are shown in Figure~\ref{histogramsign_BG13}. Although the statistical sample used here is not as large as those used in previous studies based on the PP98 model (e.g., \cite{Coburn14,Coburn15}) for the reasons explained in Section~\ref{data}, our results confirm the previously reported features of the solar wind. First, Figure~\ref{histogramsign_BG13} shows that both the histogram and the mean values (red lines) of the signed cascade rates indicate a direct cascade in the slow solar wind and an inverse cascade in the fast wind. The average cascade rates over all the statistical samples in the slow wind, $\sim 1.3\times10^{-17} J.m^{-3}.s^{-1} \sim 2.5\times10^{3} J.(kg.s)^{-1}$, are slightly higher than those reported in e.g. \cite{macbride08} ($\sim 1.9\times10^{3} J.(kg.s)^{-1}$).

\begin{figure}
\includegraphics[width=1\columnwidth]{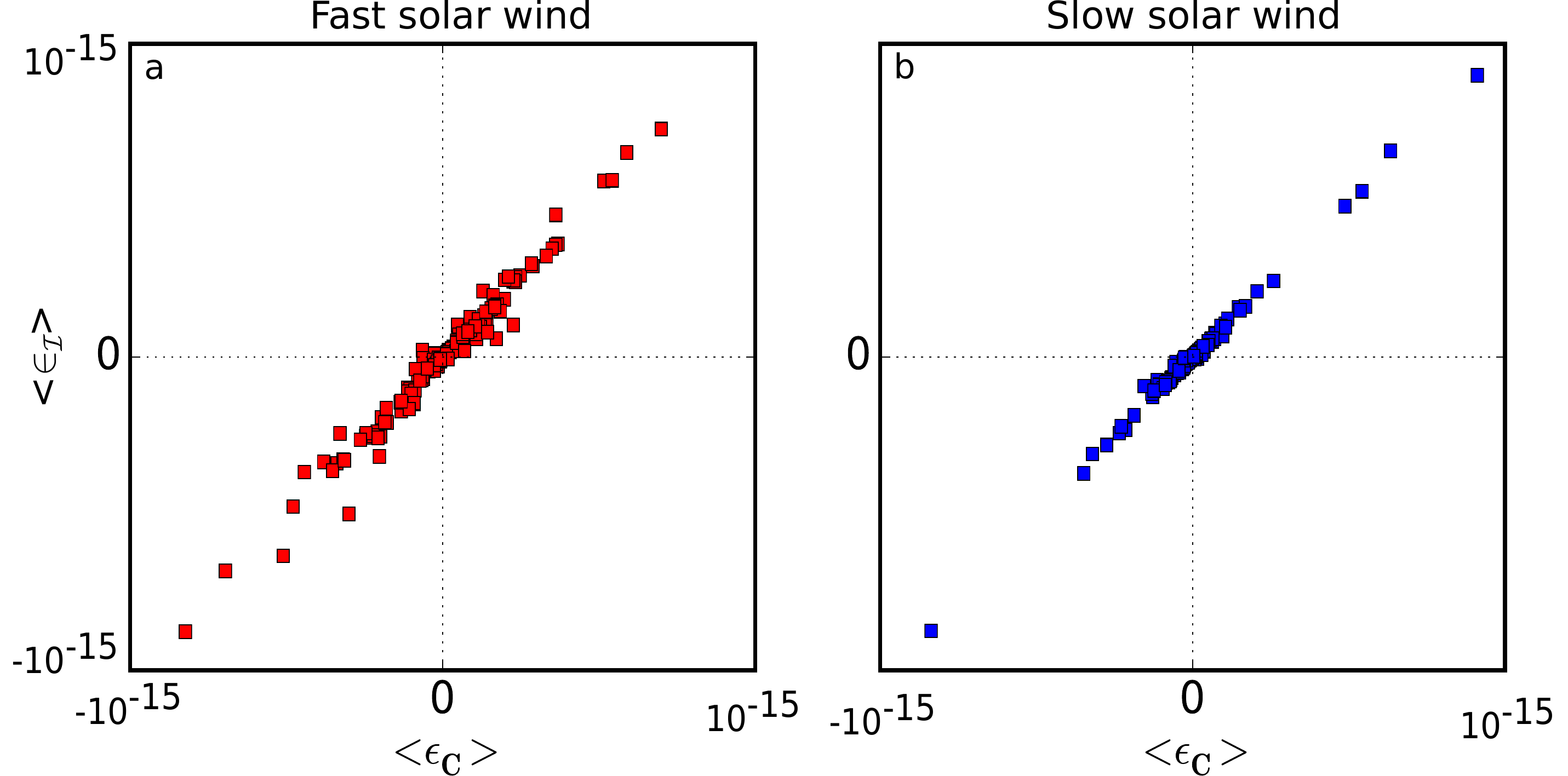}
\caption{The correlations between the estimated signed incompressible $\varepsilon_I$ and compressible $\varepsilon_C$ cascade rates 
in the fast (a) and slow (b) wind.}
\label{scattersign_BG13PP98}  
\end{figure}

The second observation is that the compressible fluctuations do not influence the direction of the cascade. This can be seen in Figure~\ref{scattersign_BG13PP98} showing the correlations between the estimated signed incompressible and compressible cascade rates $\varepsilon_I$ and $\varepsilon_C$: most of the studied cases showed the same sign for the averaged incompressible and compressible energy cascade rates.

\begin{figure}
\includegraphics[width=1\columnwidth]{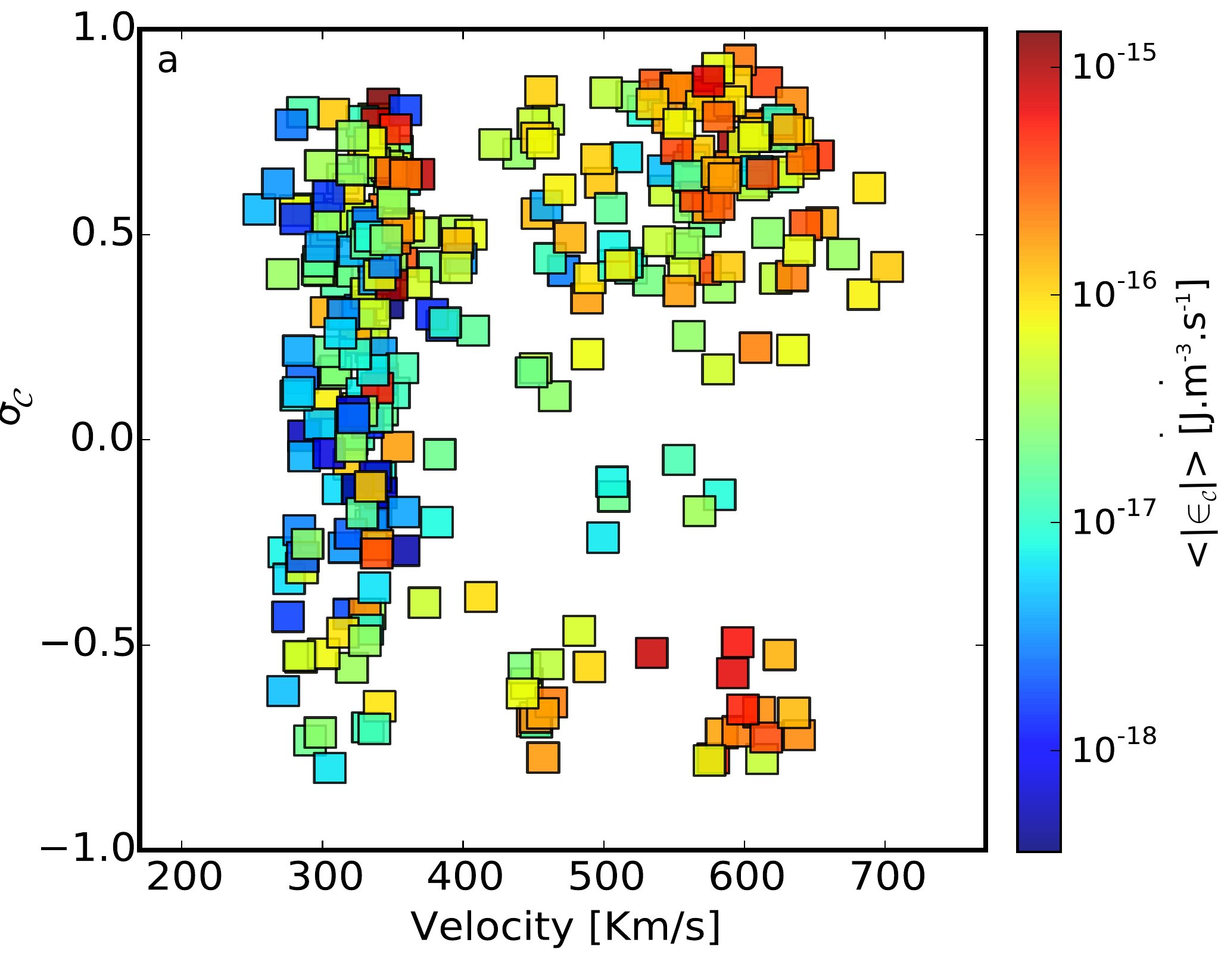}
\includegraphics[width=1\columnwidth]{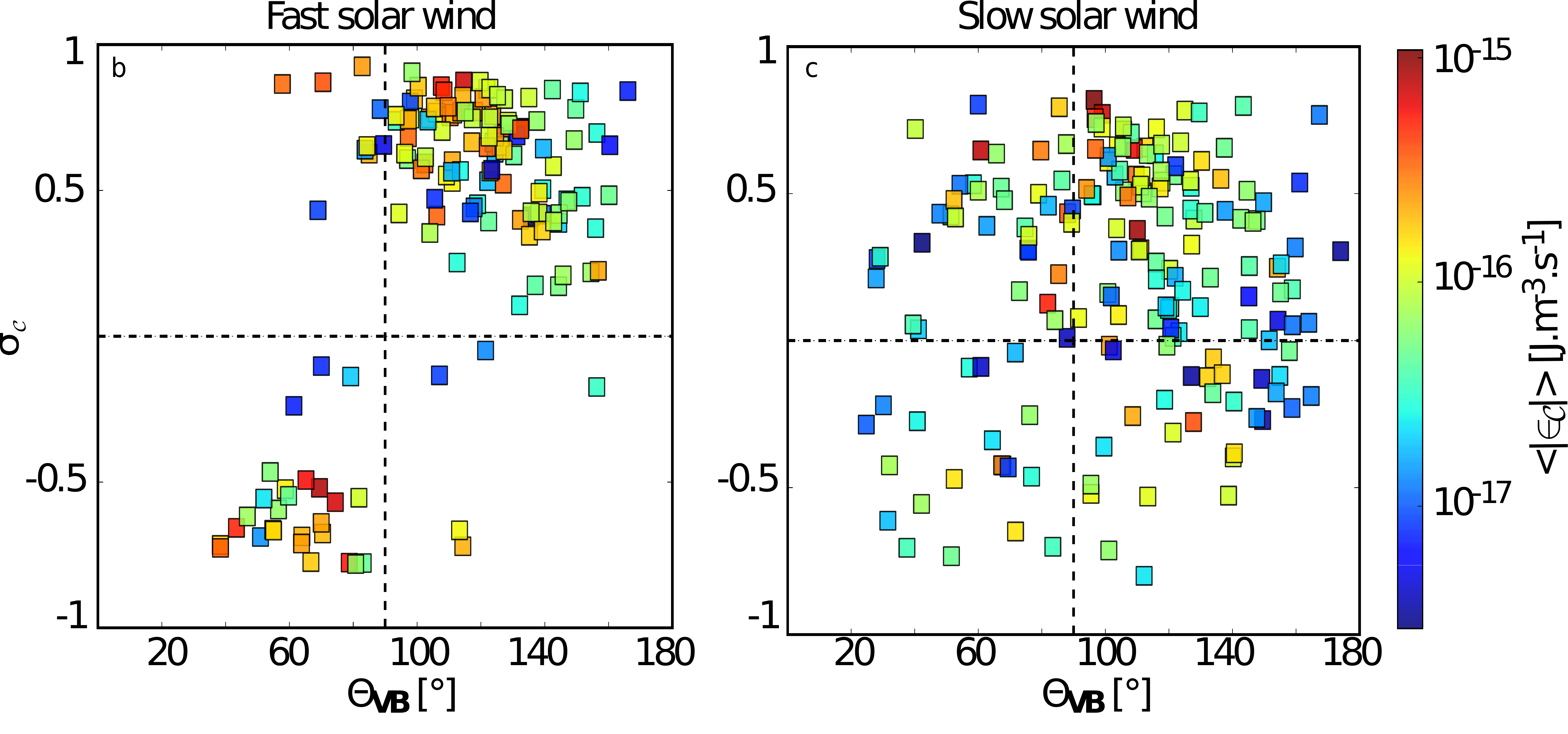}
\caption{(a) the compressible cascade rate $\varepsilon_C$ plotted as function of the cross-helicity and the solar wind speed. (b-c) the compressible cascade rate $\varepsilon_C$ plotted as function of the angle $\Theta_{\bf VB}$ and $\sigma_C$ in the slow and fast wind respectively. Outward-propagation Alfv\'en waves correspond to $\sigma_C\sim 1$ and anti-parallel to the mean magnetic field ${\bf B}_0$, while inward-propagating ones correspond to $\sigma_{C}\sim -1$ and parallel to ${\bf B}_0$.}
\label{cross_helicity}  
\end{figure}

To understand the difference in the direction of the cascade in the slow and the fast wind, we investigated the role of the cross-helicity as suggested in~\cite{Smith09}. The results of the analysis are shown in Figure~\ref{cross_helicity}. Several interesting features can be evidenced. First, we observe again the property evidenced in 
section~\ref{firstsec} that the fast wind has higher $\langle |\varepsilon_C|\rangle$ than the slow wind (Figure~\ref{cross_helicity}-(a)). Furthermore, we observe the known feature that the fast wind is generally characterized by higher values of cross-helicity $|\sigma_c|\gtrsim0.5$ with more preference for outward propagating waves ($\sigma_c>0$) (Figure~\ref{cross_helicity}-(b)). This property is not observed in the slow solar wind where $\sigma_c$ is uniformally distributed  between $\sim[-0.8,+0.8]$. Our observation of the dominance of the inverse cascade in the fast solar wind (dominated by outward propagating waves) is consistent with the finding of~\cite{Smith09} who suggested that this process could explain the survival of regions of high cross-helicity in the fast wind at large radial distances from the Sun~\citep{Roberts87}.

%-------------------------------------------------------------------------------
\subsection{Spatial anisotropy and the energy cascade rate} \label{anisot}
\begin{figure}
\includegraphics[width=1\columnwidth]{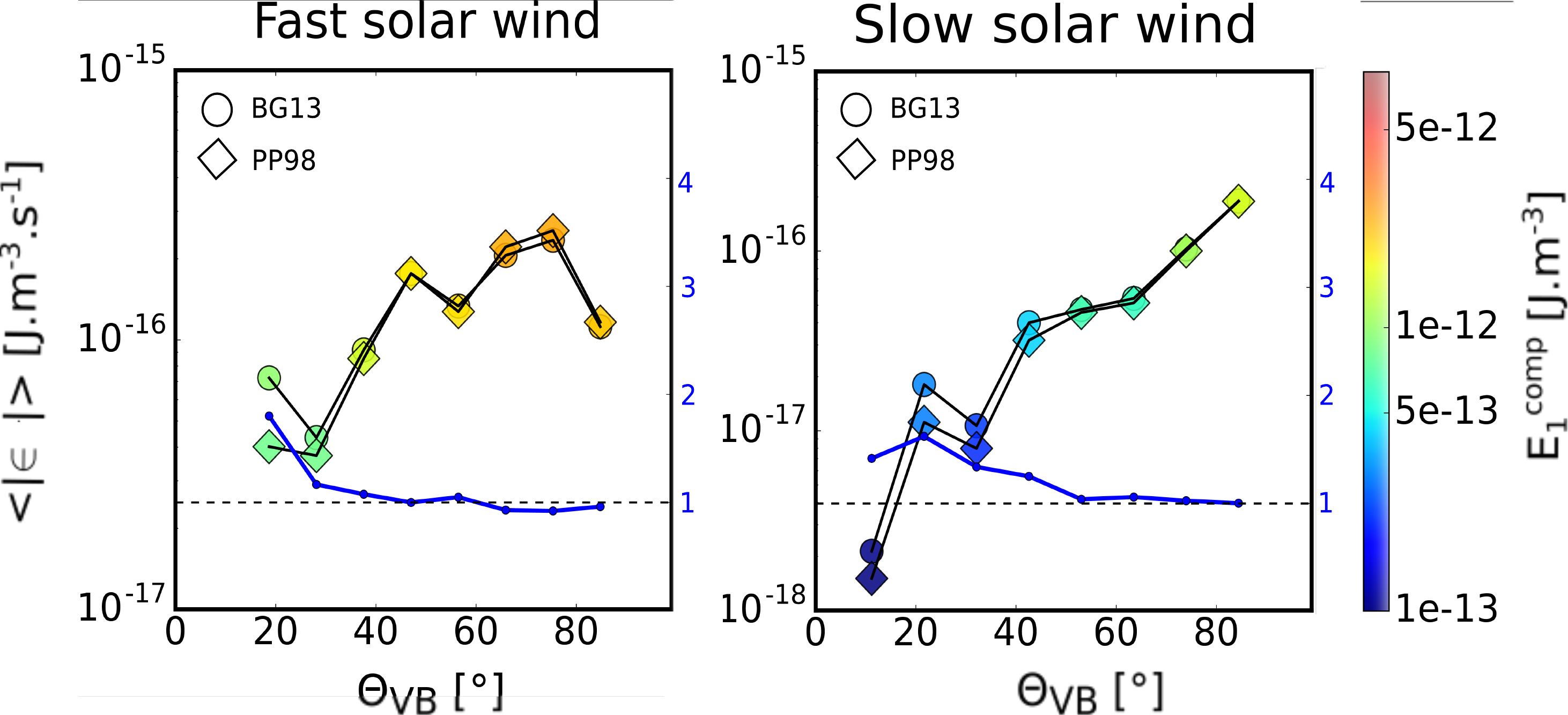}
\caption{Estimated energy cascade rates from BG13 and PP98 as a function of the angle  $\Theta_{\bf VB}$ and the total compressible energy $E_{1}^{comp}$ in the fast (Left) and slow (Right) solar wind. The blue curve represents the ratio $R=\langle|\epsilon_{C}|\rangle/\langle|\epsilon_{I}|\rangle$ as a function of $\Theta_{\bf VB}$.  }
\label{anisotropy}  
\end{figure}

%-------------------------------------------------------------------------------
In this section we explore the anisotropy nature of the cascade rate and the differences between the incompressible and compressible models. The anisotropy of the cascade rate has been previously explored using the PP98 model, and it has been shown that the cascade rate is more anisotropic in the fast than in the slow solar wind \citep{macbride08}. In the previous works, the original PP98 equations were modified to fit the limit of either 1D (slab) and 2D geometry, through the appropriate projection of the flux terms onto the two directions parallel and perpendiuclar to the mean magnetic field. Here, we do not use that approach for either the PP98 or the BG13 models. Instead, we simply examine the dependence of the estimated cascade rates on the angle $\Theta_{\bf VB}$. As we explained above, the use of the Taylor hypothesis ($\l=-V \tau$) to convert time lags $\tau$ into spatial scales implies that the analysis samples only the direction along the solar wind flow. Hence, when $\Theta_{\bf VB}\sim 0^\circ$ (resp. $\Theta_{\bf VB}\sim 90^\circ$) the analysis yields information in the direction parallel (resp. perpendicular) to the local mean magnetic field. It is worth recalling that the derivation of BG13 model does not require the isotropy assumption. 
Therefore, estimating the cascade rate using that model as function of the sampling direction of space given by the angle $\Theta_{\bf VB}$ should allow gaining insight into the anisotropic nature of the fluctuations. We used this approach by spliting our statistical samples (in the fast and slow winds) as function of the angle $\Theta_{\bf VB}$. The result is given in Figure~\ref{anisotropy}. Two important observations can be made. 
First, both models, PP98 and BG13, provide a cascade rate that is stongly depending on the angle $\Theta_{\bf VB}$. This dependence is even more pronounced in the slow wind than in the fast wind. This contrasts with the finding of \cite{macbride08} who showed no significant anisotropic cascade in the slow wind. The reason of this discrepancy may come from the criterion of uniform angle $\Theta_{\bf VB}$ used in this work, which allows us to better evidence the difference in the cascade rates parallel and perpendicularly to the mean field. However similarly to \cite{macbride08}, the
heating is smaller in the parallel direction (where $E_{1}^{comp}$ is lower) than in the perpendicular one (where $E_{1}^{comp}$ is higher) for both
winds, with a lower $\langle|\epsilon_{C}|\rangle$ for the slow compared to the fast one. 
Second, we can see that the compressible model BG13 slightly reduces the level of anisotropy in particular in the slow wind (by a factor of $R \sim 2$). This observation can easily be understood considering that, unlike the shear Alfv\'en mode in the PP98 model, the BG13 model includes also the compressible MHD (slow and fast) modes, which have a parallel magnetic field component although they are minor in the solar wind. In particular, the fast mode turbulence is shown to be isotropic from numerical simulations of MHD turbulence~\citep{Cho02}. That property naturally tends to isotropize the full turbulent fluctuations which are no longer simply guided by the mean magnetic field as in incompressible  MHD theory. 

%-----------------------------------------------------------------------
\section{DISCUSSION} \label{discussion} 
%-----------------------------------------------------------------------
Before summarizing the main finding of the present statistical study, we address some important points related to the use of compressible models to estimate the energy cascade rate in the solar wind. These points are related to the subtle role of the background (mean) density and velocity of the solar wind plasma. Other caveats will be discussed such as the role of the angle $\Theta_{\bf VB}$ and the statistical significance of the single (at given value of $\tau$) versus average (over all values of $\tau$) of the estimated cascade rates.

%========================================
\subsection{On the role of mean flow velocity}\label{dis_flow}
%========================================
In the first attempt to include compressible fluctuations in solar wind turbulence studies, \cite{Carbone09} found that the energy transfer rate $\varepsilon_{C09}$
is around $10-15$ times greater than the one given by PP98 and that amplification comes from a heuristic modification of the original (incompressible) Yaglom terms in the PP98 model. Our results showed that the compressible Yaglom term ${\cal F}_I$ does not play a significant role in enhancing $\varepsilon_C$ w.r.t. the PP98 model. The amplification comes from the new flux terms ${\cal F}_2$ and ${\cal F}_3$ that are not included in the C09 model. This discrepancy may originate from the role of the mean flow velocity that could have been erroneously included in the modified (compressible) Els\"asser variables ${\ww^\pm}$ (Equation~\ref{C09relation}) used in \cite{Carbone09}, which is much larger (by a factor $\sim 10$) than the velocity fluctuations. Indeed, when using the incompressible MHD model (PP98), the mean flow velocity is systematically suppressed in the Els\"asser variables while estimating their increments, and consequently the latter depend only on the turbulent fields fluctuations. This is consistent with the theoretical derivation of the exact laws in turbulence where a zero mean flow velocity is generally assumed. However, in the empirical compressible model of Carbone et al. 2009 (C09), the difficulty arises when dealing with the density-weighted velocity given in Equation~(\ref{eq_ww}). Because of the density dependence of the modified  Els\"asser variables, the mean flow velocity will remain involved when estimating the field increments in Equations~(\ref{C09relation}) of C09. In other words, the estimation of the cascade rate will involve not only the turbulent fluctuations but also the mean flow velocity, which is not relevant in turbulence studies and in particular for the estimate of the cascade rate. To test this hypothesis we compared the energy transfer rates computed using PP98, BG13, C09, and a modified version of the C09 model that uses the fluctuating velocity  $\vv_1$ instead of the total one (${\bf V} + \vv_1$), namely 

\begin{equation}\label{eq_ww_new}
{\bf \tilde w}^\pm = \rho^{2/3}\Big{(}\vv_1 \pm \frac{{\bf B}}{\sqrt{\rho\mu_0})}\Big{)} .
\end{equation}

\begin{figure}
\includegraphics[width=1\columnwidth]{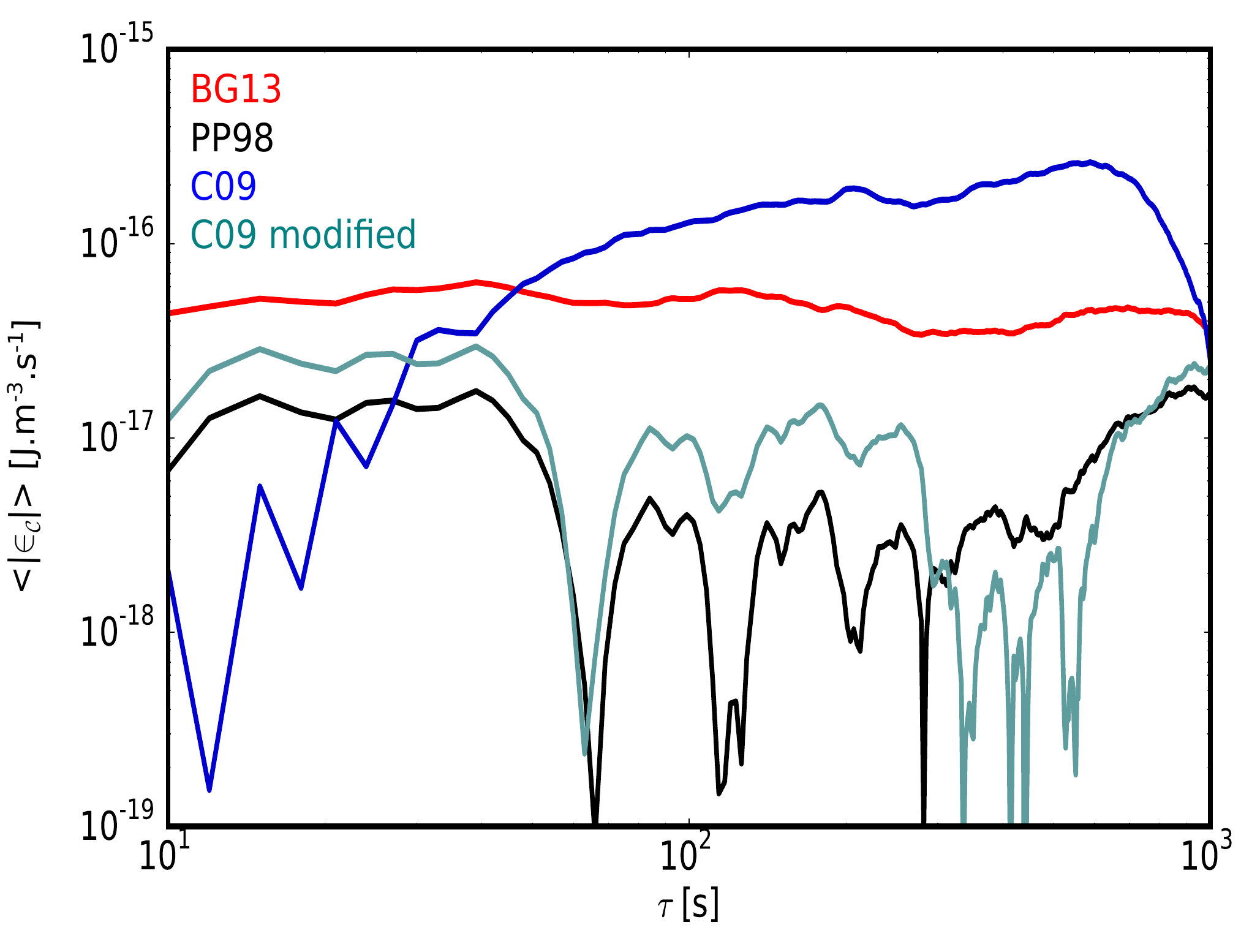}
\caption{The energy dissipation rate computed using BG13 (red), PP98 (black), C09 (blue), C09 corrected (green) for the same event of 
Figure \ref{waveforms} (On 2009-11-20 from 03:33 to 04:08).}
\label{Vtot_Carbone}  
\end{figure}

The results are shown in Figure~\ref{Vtot_Carbone}. As one can see, not only the cascade rate $\langle|\varepsilon|\rangle$ of C09 (blue) does not give a linear scaling as does the BG13 model, it also gives a cascade rate that is at least $10$ times higher than the other models. However, when using the modified C09 with the variables 
${\bf \tilde w}^\pm$, the corresponding $\langle|\varepsilon|\rangle$  (green curve) decreases and becomes comparable to the Yaglom term of PP98 (black curve). This implies that the modified C09 model, which considers compressibility corrections to the Yaglom term in the PP98 model, does not modify significantly the energy cascade rate in agreement with our finding using the BG13 model.

\begin{figure}
\includegraphics[width=1\columnwidth]{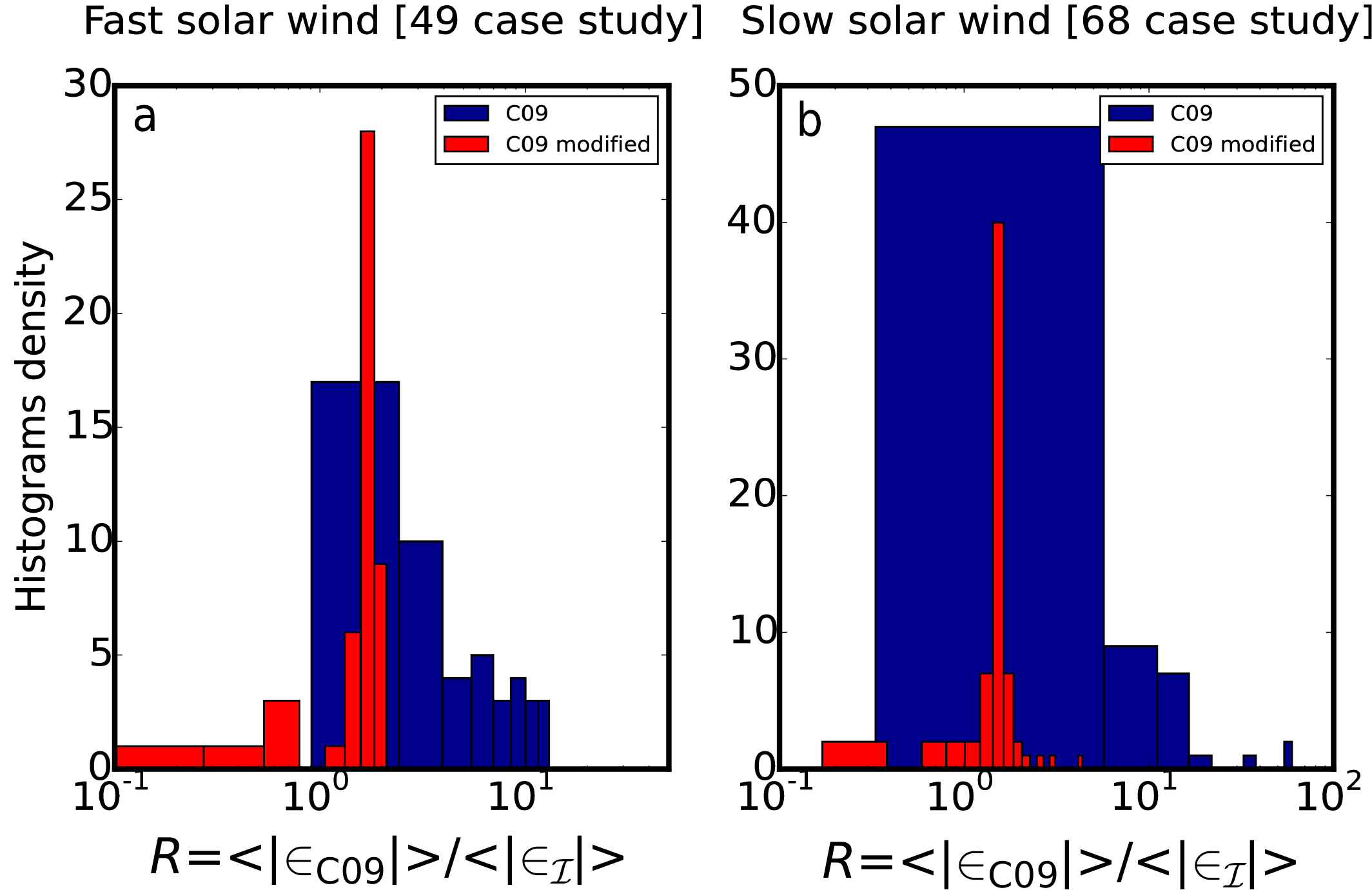}
\caption{Histograms of the ratio $R = \left\langle|\varepsilon_{C09}|\right\rangle/\left\langle|\varepsilon_{I}|\right\rangle$ using the $C09$ model (blue) 
and the corrected  one (red), in the fast (a) and the slow wind (b).}
\label{Ratio_Carbone_Mod}  
\end{figure}

This result is confirmed by a statistical analysis of all the events for 
which $\epsilon_{C09}$ is constant in sign. 
The corresponding results are shown in Figure~\ref{Ratio_Carbone_Mod}, which compares the ratios $R$ of the average energy cascade rates obtained using the original and the modified C09 models to those given by the PP98 model. As one can see, $R$ reaches values as high as $\sim 50$ both in the fast and the slow winds (blue histograms), while this ratio drops down to $\sim 1$ with the modified  $C09$ model (red histogram), in agreement with our finding using the BG13 model.

%=================================
\subsection{The mean plasma density}\label{dis_rho}
%=================================
Another point that deserves enlightment is the influence of the mean density $\rho_0$ in the BG13 model. Indeed, the original form of ${\cal{F}}_{3}$ includes the total density $\rho= \rho_{0} + \rho_{1}$ as the following~\citep{Banerjee13}:
\ba
\nabla_{\boldsymbol{\ell}}  \cdot \pmbmath{\cal F}_{3}(\ell) &=&
\nabla_{\boldsymbol{\ell}}  \cdot \left\langle 2 {\overline{\delta} \left[ \left(1 + \frac{1}{\beta} \right) e + { v_A^2  \over 2}\right] 
\delta ( \rho {\bf v})}  \right\rangle  \nonumber \\
&=& \nabla_{\boldsymbol{\ell}}  \cdot \left\langle 2 {\overline{\delta} \left[ \left(1 + \frac{1}{\beta} \right) e + 
{v_A^2  \over 2}\right] \delta ( \rho_{0} {\bf v})}  \right\rangle  \label{rho_rho1}  \\
&+& \nabla_{\boldsymbol{\ell}}  \cdot \left\langle 2 {\overline{\delta} 
\left[ \left(1 + \frac{1}{\beta} \right) e + { v_A^2  \over 2}\right] \delta ( \rho_{1} {\bf v})}  \right\rangle \nonumber  . \, \, \, \, 
\ea
In the incompressible limit ($\rho_1 \rightarrow 0$ and $\nabla \cdot  {\bf v} =0$) the divergence of $\pmbmath{\cal F}_{3}(\ell)$ vanishes. However, since in the estimation of flux terms ${\cal F}_1$, ${\cal F}_2$ and ${\cal F}_3$ using spacecraft data, we do not explicitely apply the divergence operator $\nabla_{\boldsymbol{\ell}}$, 
but rather $\nabla_{\boldsymbol{\ell}} \rightarrow 1/\ell$, it is practically impossible to ensure that ${\cal F}_3$ vanishes in the incompressible limit. To guarantee the convergence of BG13 and PP98 models in the limit of incompressibility, we kept only the second term of Equation~(\ref{rho_rho1}), while the first term can be easily transformed into source terms since
\ba
2\nabla_{\boldsymbol{\ell}}  \cdot \left\langle {\overline{\delta} X \delta ( \rho_{0} {\bf v})}  \right\rangle &=& 
2 \rho_{0} \nabla_{\boldsymbol{\ell}}  \cdot \left\langle {\overline{\delta} X \delta {\bf v}}  \right\rangle \nonumber \\
&=& \rho_{0} \nabla_{\boldsymbol{\ell}}  \cdot \left\langle X {\bf v}' - X {\bf v} + X' {\bf v}' - X' {\bf v} \right\rangle \nonumber \\
&=& \rho_{0} \nabla_{\boldsymbol{\ell}}  \cdot \left\langle X {\bf v}' - X' {\bf v} \right\rangle \nonumber \\
&=& \left\langle \rho_{0}  X \nabla'  \cdot  {\bf v}' \right\rangle  + \left\langle  \rho_{0} X' \nabla  \cdot {\bf v} \right\rangle , \label{divrel}
\ea
where $X=\overline{\delta} \left[ \left(1 + \frac{1}{\beta} \right) e + {v_A^2  \over 2}\right]$. It is easy to see that both expressions (\ref{divrel}) and the flux term $\pmbmath{\cal F}_{3}(\ell)$ of Equation~\ref{fcphi} converge to zero in the incompressible limit (i.e., $\nabla  \cdot {\bf v}=0$ and $\rho_1$ = 0).

%==============================================
\subsection{The influence of the angle $\Theta_{\bf VB}$}\label{dis_theta}
%==============================================
In Section~\ref{data} we emphasized the importance of having relatively stationary angles $\Theta_{\bf VB}$ in order to have a more reliable estimate of the energy cascade rate (both its sign and its absolute value) when dealing with single spacecraft data, and regardless of the theoretical model used. Here we discuss two possible effects of the non stationarity of the angle $\Theta_{\bf VB}$ that may influence the estimation of the cascade rate.  

\begin{figure}
\includegraphics[width=1\columnwidth]{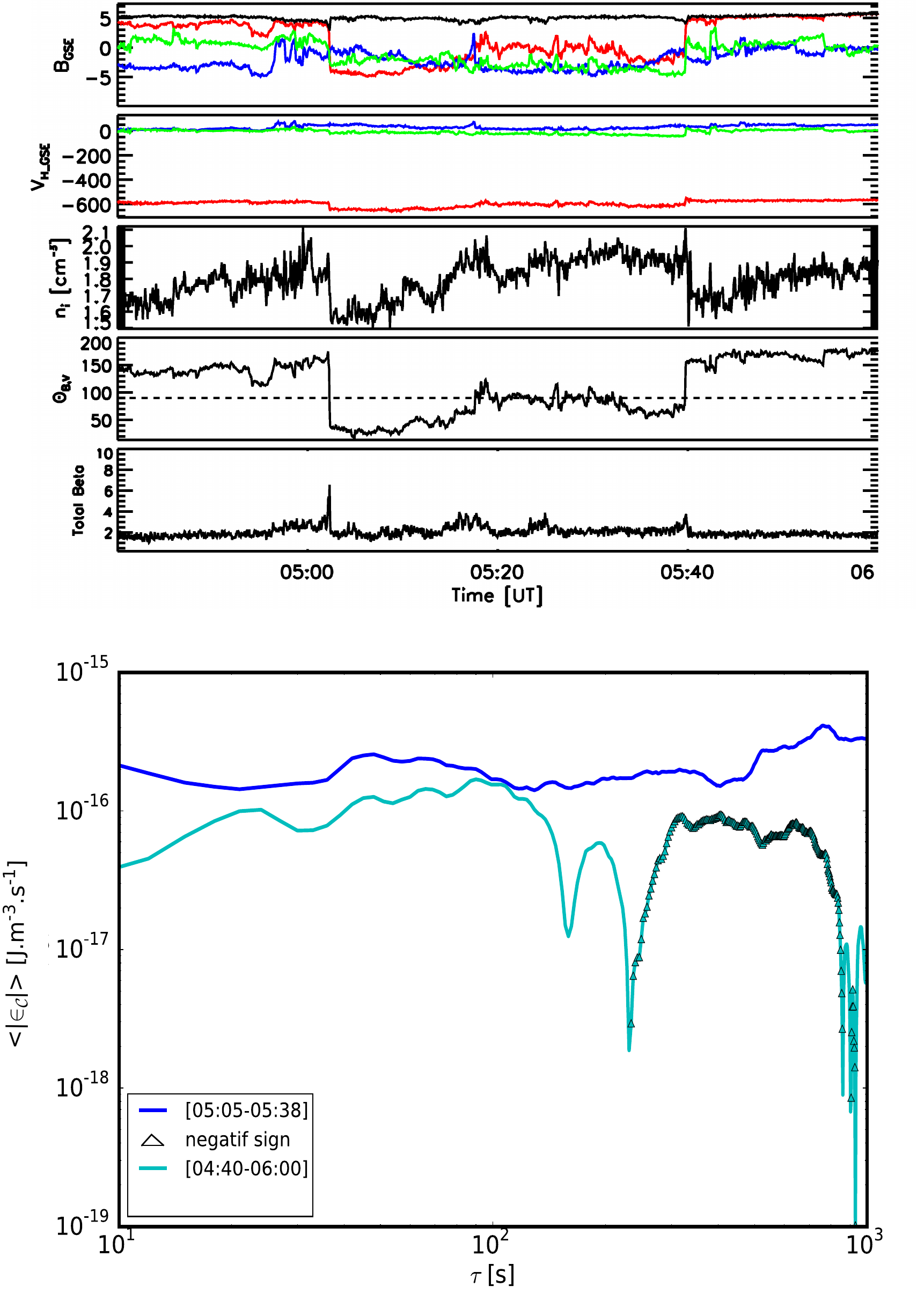}
\caption{Top: sample of the fast solar wind data with sharp variations of $\Theta_{\bf VB}$ on 2010-12-14 from 04:40 to 06:00. 
Bottom: cascade rates computed for the entire signal (green) and after filtering (blue) to exclude the $\Theta_{\bf VB}$ rotation.}
\label{AngleRotationF}  
\end{figure}

Let us first start with the case of the presence of sharp variations (i.e., discontinuities) in the angle $\Theta_{\bf VB}$ as in the example of Figure~\ref{AngleRotationF}. Such discontinuities may be due to different reasons such as the crossings of strong current sheets frequently observed in the solar wind and the magnetosheath~\cite{Gosling08,Chasapis15}. We estimated the energy cascade rate using BG13 from a long but non stationary time interval (04:40-06:00) that contained two discontinuities in $\Theta_{\bf VB}$ (about 05:00 and 05:40) and from shorter one (05:05-05:38) where such discontinuities were excluded. The results are shown in Figure~\ref{AngleRotationF} (bottom). As one can see the long non stationary time interval yields a non uniform energy cascade rate which changes its sign, whereas the shorter one where the $\Theta_{\bf VB}$ sharp discontinuities were excluded is more uniform and has a constant sign. This result should balance the usual wisdom arguing to use long time intervals (i.e., large number of data points) to guarantee the statistical convergence of the third-order moments estimates~(e.g., \cite{Podesta09}): the existence of a very few (i.e. statistically minor) sharp discontinuities as those in Figure~\ref{AngleRotationF} can significantly influence the estimates of the cascade rate as we showed here.

The second possible effect of the angle $\Theta_{\bf VB}$ can come from its steady but significant variation in a single time interval. Indeed, as we argued in Section~\ref{data}, the Taylor frozen-in-flow assumption generally used on single spacecraft data allows one to convert the time sampling of the data into a 1D spatial sampling of the turbulent fluctuations along the flow direction. In anisotropic turbulence, the direction of the spatial sampling carries therefore a particular importance since the sampling can be either parallel ($\Theta_{\bf VB}\sim 0^\circ$) or perpendicular ($\Theta_{\bf VB}\sim 90^\circ$) to the mean field. These two directions, as demonstated in Figure~\ref{anisotropy}, have different values of the energy cascade rate. Therefore, if $\Theta_{\bf VB}$ oscillates strongly between $0^\circ$ and $90^\circ$ then the analysis would mix between the two cascade rates estimated along the direction parallel and perpendicular to the local magnetic field, and would lead to higher uncertainty in the estimated values. This might be the reason that explains the discrepancy in the cascade rate in the slow solar wind found between our results and those of \cite{macbride08}.

%==============================================
\subsection{Mean value of cascade rate and sign change}\label{dis_sign}
%==============================================
As explained in Section~\ref{data}, among the criteria that we used to select our statistical samples is the constant sign of the estimated cascade rate $\varepsilon_C$ over the time lag $\tau \in[10,1000]$s. This step is necessary in order to get reliable estimate of the mean cascade rate $\langle \varepsilon_C \rangle$ averaged over all the time lags $\tau$. Indeed, if the sign of $\varepsilon$ changes, the resulting average will yield (by cancellation) lower values of the cascade rates. Another alternative to this approach has been used in previous works based on performing statistical studies of the cascade rate obtained at a given value of the time lag $\tau$~\citep{macbride08,Smith09}. The choice of the particular $\tau$ value has not been justified apart from the fact that it belongs to the inertial range. The drawback of this approach is that, since the sign of $\varepsilon$ can vary within the inertial range as can be seen in Figure~\ref{AngleRotationF} and in e.g. in \cite{luca}, the choice of the value of $\tau$ may influence the conclusion regarding the nature (direct versus inverse) of the turbulent cascade. 

\begin{figure}
\includegraphics[width=1\columnwidth]{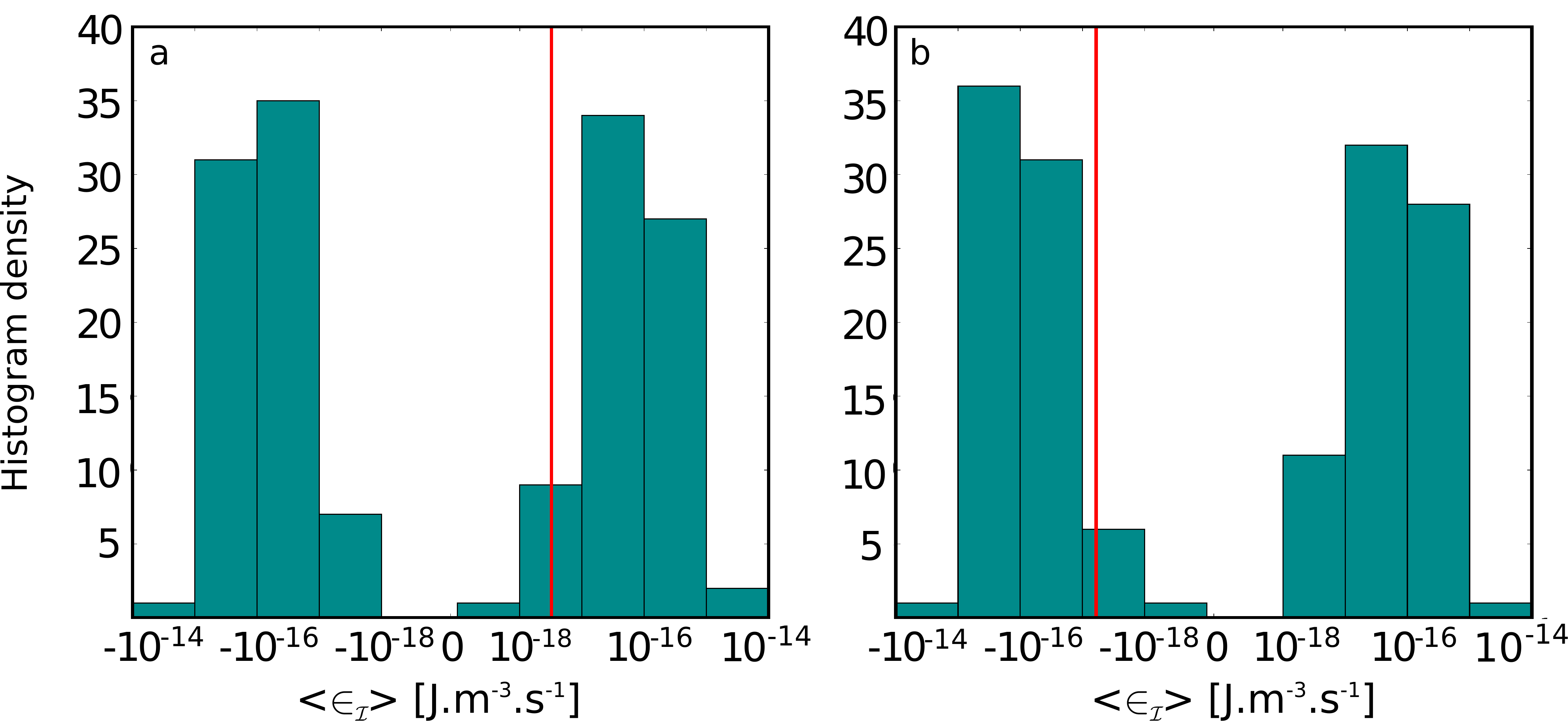}
\caption{Histogram of the estimated cascade rate $\varepsilon_I$ in the 
fast wind from PP98 at two different values of the time lag, (a) $\tau = 21$ s and (b) $\tau=81$ s.
The red line represents the mean value of $\varepsilon_{I}$ for the given value of $\tau$.}
\label{dis_tau}  
\end{figure}

Figure~\ref{dis_tau} shows the histogram of $\varepsilon_{I}$ computed using PP98 at different values of $\tau$. For $\tau = 21$ s (Figure~\ref{dis_tau}-(a)) $\langle \varepsilon_{I} \rangle$ is positive, implying a direct cascade, whereas for $\tau = 81$ s (Figure~\ref{dis_tau}-(b)), $\langle \varepsilon_{I} \rangle$ is negative indicating an inverse cascade. This result underlines the need to be cautious when interpreting statistical results about cascade rates estimated at a single value of the time lag $\tau$ even when it belongs to the inertial range. 

%-----------------------------------------------------------------------
\section{SUMMARY AND CONCLUSIONS} \label{conclusion} 
%-----------------------------------------------------------------------
In this paper we provided the first statistical study of the compressible energy cascade rate in fast and slow solar wind MHD turbulence using a large survey of the THEMIS/ARTEMIS spacecraft data. The work is based on the reduced form of the  isothermal compressible MHD turbulence model recently derived in~\cite{Banerjee13}. Several new results have been obtained, which include the amplification of the cascade rate and its slight isotropization (in particular in the slow wind) due to compressible fluctuations and a better definition of the inertial range thanks to a steadier (in value and sign) of the estimated compressible cascade rate over more than two decades of scales in comparison to the incompressible PP98 model. The new flux terms contained in the BG13 model were shown to play a leading role in amplifying the compressible energy cascade rate rather than the modified compressible Yaglom term. This result desagrees with the finding of \cite{Carbone09} who used an heurtistic compressible model based on a modification of the Yaglom term in PP98 model via density fluctuations. That discrepancy motivated a comparative study with the C09 model, which eventually showed that the origin of the cascade rate amplification found in \cite{Carbone09} is due to the mean solar wind velocity included in that estimation through the modified (compressible) Els\"asser variables. Other important results have been obtained such as the new empirical scaling laws relating the new compressible cacade rate to the sonic turbulent Mach number, and to the different components (magnetic, kinetic and internal) of the fluctuating energy. Interpreting those empirical laws requires further theoretical investigations. Several caveats related to the data selection and to the role angle $\Theta_{\bf VB}$ on the convergence of the energy cascade rate were highlighted. 

While this works based on the new BG13 model undoubtfully sheds light onto new features of solar wind turbulence, it remains however a perfectible model. Two particular aspects require to be improved. The first one is related to the source terms that could not have been estimated in this work using single spacecraft data (as they involve local divergences of the Alfv\'en and the plasma velocity fluctuations). Reliable estimation of those terms can be done using  multispacecraft observations. Cluster spacecraft offer that possibility but the plasma data (density, velocity and temperature) are available only on two (out of four) spacecraft which does not allow us to obtain 3D estimation of the source terms. The recently launched MMS mission offers a more interesting alternative as both the magnetic field and plasma data are available on the four spacecraft. However, the mission in its current phase explores only the magnetopause and the magnetosheath regions (with a focus on the former) and will reach out in the solar wind only in 2018. Another possible shortcoming is the spacecraft separation ($\sim 10$km), which would not allow accurate estimation of the gradients at scales of the inertial range $>100$km)~\citep{Robert98}. Other than spacecraft data, numerical simulation of isothermal compressible MHD  turbulence should allow for straightforward  estimation of the source terms and their comparison to the flux terms. This task is planned for the upcoming months. On a longer run, the BG13 model needs to be extended to more general closure equations such as the polytropic one to go beyond the current simplified isothermal closure. 
\paragraph*{Acknowledgment.}
The THEMIS/ARTEMIS data come from the AMDA data base (http://amda.cdpp.eu/).
We are grateful to Dr. O. Le Contel and Dr S. Banerjee for useful discussions. 
FS acknowledges financial support from the ANR project THESOW, grant ANR-11-JS56-0008. The french participation in the THEMIS/ARTEMIS mission is funded by CNES and CNRS.

\end{document}